\documentclass[amsmath, amsthm, amssymb,  superscriptaddress, onecolumn, 10pt]{revtex4-2}
\usepackage[nearskip,margin = 0pt,caption=false]{subfig}
\usepackage{siunitx}
\usepackage[hidelinks]{hyperref}
\usepackage{tikz}

\usepackage[labelfont=bf, justification=raggedright, singlelinecheck=false, figurename={Fig.}, tablename={Table}, labelsep=period, font=small]{caption}

\newcommand{\figlabel}[1]{\textbf{\MakeLowercase{#1}},}

\begin{document}
\title{Ultrafast one-chip optical receiver with functional metasurface}
\author{Go~Soma}
\email{go.soma@tlab.t.u-tokyo.ac.jp}
\affiliation{School of Engineering, The University of Tokyo, Tokyo 113-8656, Japan}
\author{Tomohiro~Akazawa}
\affiliation{School of Engineering, The University of Tokyo, Tokyo 113-8656, Japan}
\author{Eisaku~Kato}
\affiliation{School of Engineering, The University of Tokyo, Tokyo 113-8656, Japan}
\author{Kento~Komatsu}
\affiliation{School of Engineering, The University of Tokyo, Tokyo 113-8656, Japan}
\affiliation{Present address: Sumitomo Electric Industries, Ltd., Kanagawa 244-8588, Japan}
\author{Mitsuru~Takenaka}
\affiliation{School of Engineering, The University of Tokyo, Tokyo 113-8656, Japan}
\author{Yoshiaki~Nakano}
\affiliation{School of Engineering, The University of Tokyo, Tokyo 113-8656, Japan}
\author{Takuo~Tanemura}
\email{tanemura@ee.t.u-tokyo.ac.jp}
\affiliation{School of Engineering, The University of Tokyo, Tokyo 113-8656, Japan}

\begin{abstract}
\noindent
High-speed optical receivers are crucial in modern optical communication systems. While complex photonic integrated circuits (PICs) are widely employed to harness the full degrees of freedom (DOFs) of light for efficient data transmission, their waveguide nature inherently constrains two-dimensional spatial scaling to accommodate a large number of optical signals in parallel. Here, we present a novel optical receiver platform that fully exploits the high spatial parallelism and ultrabroad bandwidth of light, while leveraging all DOFs – intensity, phase, and polarization. Our solution integrates a thin metasurface, composed of silicon nanoposts, with ultrafast membrane photodetectors on a compact chip. The metasurface provides all the functionalities of conventional PICs for normal-incident spatially parallelized light, enabling high-speed detection of optical signals in various modulation formats, including simultaneous detection of 320-gigabits-per-second four-channel signals and coherent detection of a 240-gigabits-per-second signal.
\end{abstract}

\maketitle

\noindent
Metasurfaces (MSs) are two-dimensional (2D) arrays of subwavelength nanostructures that can manipulate the intensity, phase, and polarization of transmitted light through ultrathin flat elements, offering a compact alternative to conventional bulky optical systems \cite{Yu2011-ey, Khorasaninejad2017-sb, Arbabi2022-jk, Gu2022-ky, Kuznetsov2024-lv, Ha2024-mf}. 
With judiciously designed MSs, various functional devices have been realized, including achromatic lenses \cite{Chen2018-lu, Shrestha2018-cx, Wang2018-cm, Chen2019-eo, Hu2023-ni}, polarimeters \cite{Rubin2019-tb, Arbabi2018-vz, Pors2015-il, Yang2018-yh, Shah2022-fm, Soma2024-fc}, color routers \cite{Faraji-Dana2018-xc, Zhu2019-xe, Miyata2021-cg, Zou2022-xp}, holograms \cite{Ni2013-be, Zheng2015-xm, Wang2016-aj, Arbabi2015-rv, Balthasar_Mueller2017-ex, Ren2020-bs, Bao2022-mz, Zaidi2024-lk}, and augmented/virtual reality and display devices \cite{Lee2018-eq, Joo2020-lp, Li2021-gi, Li2022-zu, neshev2023enabling, Gopakumar2024-ad}.
While these imaging, sensing, and display applications have demonstrated the remarkable potential of MSs over the past decade, modern high-speed optical communication systems represent another promising frontier for leveraging the rich capabilities of MSs.

For instance, coherent optical transmission systems have enabled long-haul, high-capacity data transport by utilizing the full degrees of freedom (DOFs) of light -- intensity, phase, and polarization -- to encode information. Additionally, the space-division multiplexing (SDM) scheme, which employs spatially parallelized channels within multi-core and multi-mode fibers (MCFs/MMFs), is envisioned as the promising next-generation technology to further scale transmission capacity \cite{Richardson2013-qw, Mizuno2016-na, puttnam2021space}. With the evolution of these paradigms, optical transceivers have become increasingly complex. A coherent receiver (CR), for example, requires multiple high-speed photodetectors (PDs), a polarization beam splitter (PBS), and precisely phase-controlled optical hybrids to retrieve dual-polarization in-phase and quadrature (IQ) components of optical signals \cite{Kikuchi2016-pd}. 
However, today, these transceivers are implemented on waveguide-based photonic integrated circuits (PICs) \cite{Dong2014-qf,  Ogiso2017-hb, Yagi2018-do, Porto2022-nz, Xu2020-pm}, which are not easily scalable to a 2D array to accommodate a large number of spatial channels efficiently \cite{Marchetti2019-pv}.
Although some attempts have been reported, demonstrating discrete MS-based devices for optical communication \cite{Oh2022-tp, Oh2024-ux, Soma2023-fx, Komatsu2024-pi, Komatsu2024-OFC, Ren2024-wz}, MS-enabled fully integrated high-speed optical transceivers have not been realized to the best of our knowledge.

Here, we present a novel high-speed optical receiver operating at the 1550-nm telecommunication wavelength that employs a functional dielectric MS integrated with an ultrafast membrane indium-gallium-arsenide (InGaAs) PD array (PDA). 
A micrometer-thick MS, composed of silicon (Si) nanoposts, provides all the necessary operations on normally incident light, offering functionalities equivalent to conventional millimeter-square-scale waveguide-based PICs.
Combined with the membrane InGaAs PD that can efficiently detect infrared light within a submicrometer thickness, we realize various types of ultrafast receivers on a compact chip that can accept spatially parallelized, normally incident optical signals directly from a multi-channel fiber without using bulky optics. 
High-speed signals in various formats, such as 240-Gbit/s 64-ary quadrature amplitude modulation (64QAM) and 320-Gbit/s four-channel four-level pulse amplitude modulation (PAM4), are successfully demodulated with a bit error rate (BER) of less than $8.8\times 10^{-3}$, well below the soft-decision forward error correction (SD-FEC) threshold.

\begin{figure}[tb]
\centering\includegraphics{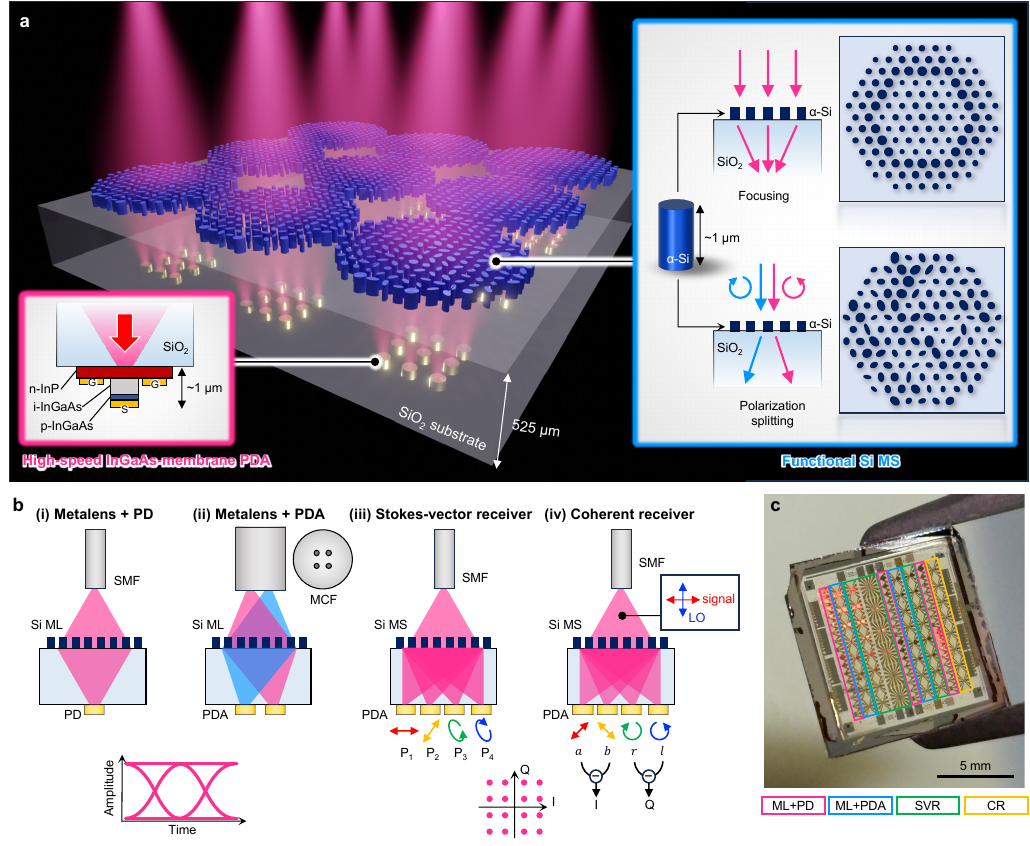}
\caption{
\textbf{One-chip optical receiver platform with an integrated Si MS and membrane InGaAs PDA.} 
\figlabel{a} Schematic illustration of our receiver platform, where ultrathin MS and high-speed membrane InGaAs PD layers are integrated on both sides of a transparent SiO$_2$ substrate. Spatially parallelized input signals are incident from the MS side and focused onto arrayed PDs. The MS offers various advanced functionalities, including focusing, splitting, and polarization manipulation, as shown in the right inset. The left inset shows the InGaAs/InP p-i-n structure of the high-speed membrane PD, which is directly bonded on the other side of the substrate. 
\figlabel{b} Schematics of four types of receivers demonstrated in this work. (i) Single-channel ML-integrated PD. (ii) ML-integrated PDA to detect parallel signals from an MCF. (iii) SVR with an integrated MS that sorts input light to four different polarization bases (P$_1$, P$_2$, P$_3$, and P$_4$) and focuses them on a four-channel PDA. (iv) CR with an integrated MS that splits input light to four polarization states ($a$, $b$, $r$, and $l$) and focuses them on a four-channel PDA. The signal and LO light from an SMF are incident on the MS with $x$ and $y$ orthogonal polarizations, so that IQ components of the signal can be retrieved from the four photocurrents.
\figlabel{c} Photograph of the receiver chip fabricated on a 1.2-cm-squared SiO$_2$ substrate, which contains 94~receivers with four different configurations.
}
\label{fig:platform}
\end{figure}
\section*{Device concept and fabrication}
\label{sec:intro}
\noindent
Figure \ref{fig:platform}a shows the concept of our one-chip receiver platform, which comprises membrane InGaAs PDA and functional Si MS layers integrated on both sides of a 525-{\textmu}m-thick fused silica (SiO$_2$) substrate.
The input signal light is normally incident on the MS side, which consists of a densely located array of 1050-nm-high Si nanoposts, operating as meta-atoms. 
The light transmitted through the MS layer is received by the membrane PDA, which consists of vertical p-i-n diodes with a 500-nm-thick i-InGaAs absorption layer sandwiched by p/n-doped InGaAs and indium phosphide (InP) layers (see Extended Data Fig.~\ref{fig:fabrication}a for the detailed profile).

The PDA and MS layers of our platform enable unique properties unattainable with conventional surface-normal receivers.
First, the membrane InGaAs/InP p-i-n PD provides efficient opto-electric (O-E) conversion in the 1550-nm wavelength band with ultralow capacitance and high electrical conductivity, resulting in ultrahigh O-E bandwidth exceeding 100 GHz \cite{Nozaki2016-yj, Shen2016-ve}.  
Second, the Si MS layer offers versatile functionalities beyond those of a simple focusing lens. 
By carefully designing the geometries of Si nanoposts, both the wavefront and polarization state of transmitted light can be controlled to enable advanced functionalities. These include beam splitting, polarization conversion, and demultiplexing, in addition to the basic functionality of a meta-lens (ML) to focus incident light onto a specific PD.
We can, for example, design an MS to demultiplex multiple optical signals from an MCF and focus to different PDs (Fig.~\ref{fig:platform}b-(ii)).
Moreover, similar to PICs, complex optical components, such as a PBS and optical hybrid required in Stokes-vector receivers (SVRs) and CRs, can be realized in a surface-normal configuration (Fig.~\ref{fig:platform}b-(iii), (iv)). 
The proposed platform, therefore, enables one-chip ultrafast optical receivers for normally incident SDM signals in various modulation formats without requiring additional optical components.

The one-chip receivers were fabricated by first wafer-bonding an InGaAs/InP membrane PD layer onto a SiO$_2$ substrate. After the PD structures were fabricated,  MS patterns were formed on an amorphous Si ($\alpha$-Si) layer deposited on the other side of the substrate through electron-beam lithography and reactive-ion etching processes (detailed fabrication processes are provided in Methods and Extended Data Fig.~\ref{fig:fabrication}b).

\section*{Experimental results}
Figure~\ref{fig:platform}c shows the fabricated chip with 1.2-cm-squared size, which contains 94 receivers in total with four different configurations: (i) single-channel ML-integrated intensity-modulation and direct-detection (IM-DD) receiver, (ii) four-channel ML-integrated IM-DD receivers to detect spatially parallelized signals from an MCF, (iii) SVR with a polarization-sorting MS, and (iv) CR with a polarization-splitting MS operating as an optical hybrid.
The chip was mounted on a stage and characterized using radio-frequency (RF) probes (details of design and measurement for each device are provided in Methods and Exteded Data Fig.~\ref{fig:setup}).

\begin{figure}[tb!]
\centering\includegraphics{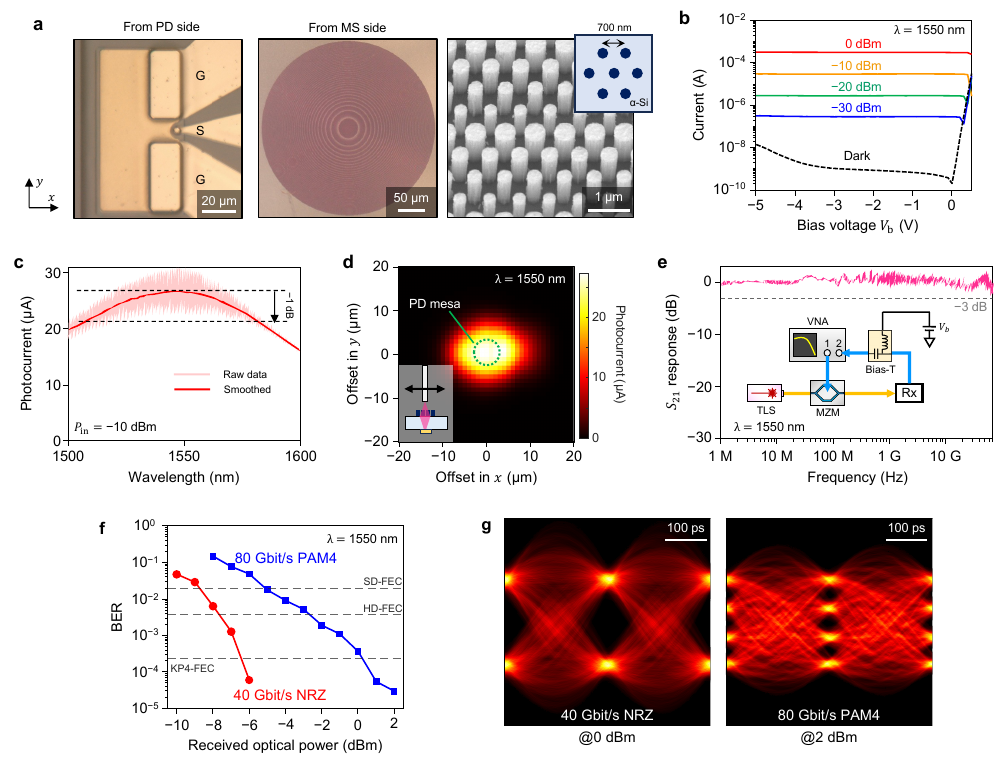}
\caption{
\textbf{Results of an ML-integrated IM-DD receiver.}
\figlabel{a} Optical microscope and SEM images of the fabricated device, observed from the PD and MS sides of the chip. 
The diameter of the circular PD mesa is $D_\mathrm{PD}=6$~\textmu m. 
The MS comprises $\mathrm{\alpha}$-Si cylindrical nanoposts arranged on a triangular lattice with a lattice constant of 700~nm (inset).
\figlabel{b} I-V characteristics of the PD measured under various input optical power ($P_\mathrm{in}$) at 1550-nm wavelength. 
\figlabel{c} Photocurrent spectrum measured at $P_\mathrm{in}=-10$~dBm. The dark red line represents a smoothed curve of the raw data (bright red line). The oscillations are attributed to the Fabry-Pérot resonance between the output facet of the SMF and the SiO$_2$-InP interface at the membrane PD.
\figlabel{d} Dependence of receiver sensitivity on the in-plane position of the input SMF, measured at 1550-nm wavelength with $P_\mathrm{in} = –10$~dBm. The green dotted line shows the PD mesa with a diameter ($D_\mathrm{PD}$) of 6~\textmu m.
\figlabel{e} Frequency response measured at a bias voltage of $V_b = -3$~V. The inset shows the schematic of the measurement setup. TLS:~tunable laser source. MZM:~Mach-Zehnder modulator. VNA:~vector network analyzer. Rx:~receiver. 
\figlabel{f} Measured BERs of 40-Gbit/s NRZ and 80-Gbit/s PAM4 signals at 1550-nm wavelength as a function of the received optical power. BER thresholds of various FEC formats are also plotted as a reference.
\figlabel{g} Eye diagrams of received 40-Gbit/s NRZ and 80-Gbit/s PAM4 signals.
}
\label{fig:ML-PD}
\end{figure}

\subsection*{ML-integrated single-channel IM-DD receiver}
\noindent
Figure~\ref{fig:ML-PD} shows the first example of a single-channel IM-DD receiver, where a 340-{\textmu}m-diameter MS, designed to work as an ML, is integrated on the other side of a 6-{\textmu}m-diameter InGaAs/InP membrane PD.
The MS is composed of $\mathrm{\alpha}$-Si nanoposts arranged on a triangular lattice with a lattice constant of 700~nm (Fig.~\ref{fig:ML-PD}a, inset). They are designed to focus a beam emitted from a single-mode fiber (SMF) onto the PD. 
The distance between the output facet of the SMF and the MS is set to 841~{\textmu}m, so the mode field diameter (MFD) of the beam incident to the MS is around 160 {\textmu}m.

We first calculated the optical phase shift $\varphi_\mathrm{ML}(x,y)$ required at each position $(x,y)$ on the MS to achieve the desired focusing functionality.
The diameter of each cylindrical $\mathrm{\alpha}$-Si nanopost was then determined to obtain $\varphi_\mathrm{ML}(x,y)$ with a minimal insertion loss.
Figure~\ref{fig:ML-PD}a shows optical microscope and scanning electron microscope (SEM) images of the fabricated device, observed from the MS and PD sides of the chip.
Ground-signal-ground (GSG) electrode pads are integrated with the PD to enable high-speed characterization.

The current-voltage (I-V) characteristic is measured under different values of input optical power, $P_\mathrm{in}$, at a wavelength of 1550~nm (Fig.~\ref{fig:ML-PD}b). The dark current is around 1~nA at a bias voltage of $V_b=-3$~V.  
From these results, the overall receiver sensitivity, which includes the focusing efficiency of ML, is derived to be 0.27~A/W. Using the simulated absorption efficiency of around 30\% by the membrane PD, the focusing efficiency of the fabricated ML is estimated to be around 70\%. 
From the measured photocurrent spectrum (Fig.~\ref{fig:ML-PD}c), we can confirm that the device operates over a wide wavelength range with 1-dB and 3-dB bandwidths of 73~nm (1508 $\sim$ 1581~nm) and over 100~nm, respectively. 
To examine the focusing functionality of the ML, the measured photocurrent is plotted as a function of the in-plane position of the input SMF in Fig.~\ref{fig:ML-PD}d.
From this plot, it is confirmed that the sensitivity drops rapidly as we move the SMF by $\pm$10~{\textmu}m in both $x$ and $y$ directions.
Since this sensitivity distribution should represent the convolutional integral of the incident beam profile at the PD plane and the aperture of the PD, having a mesa diameter of 6 {\textmu}m, we can conclude that the fabricated MS effectively focuses the input beam with an MFD of around 160 {\textmu}m down to around 15~{\textmu}m at the input of the membrane PD.

Figure~\ref{fig:ML-PD}e shows the O-E frequency response of the device, measured by a vector network analyzer (VNA). 
The 3-dB bandwidth is more than 70~GHz, which was limited by the maximum frequency range of our VNA.
Finally, Figs.~\ref{fig:ML-PD}f,g show the results of the high-speed signal detection experiment.
Although the fabricated device exhibited an ultrabroad bandwidth exceeding 70~GHz, we could only employ 40-Gbaud signals due to the bandwidth constraint of the real-time oscilloscope available at the time of the experiment.
For both 40-Gbit/s non-return-to-zero (NRZ) and 80-Gbit/s PAM4 signals, BERs below the KP4-FEC threshold of 2.4$\times 10^{-4}$  are obtained with clear eye openings. In addition, wavelength-insensitive operation across the entire $C$ band (1530 $\sim$ 1570~nm) is confirmed (Extended Data Fig.~\ref{fig:IMDD-C}).

\subsection*{ML-integrated four-channel IM-DD receiver}

\begin{figure}[bt]
\centering\includegraphics{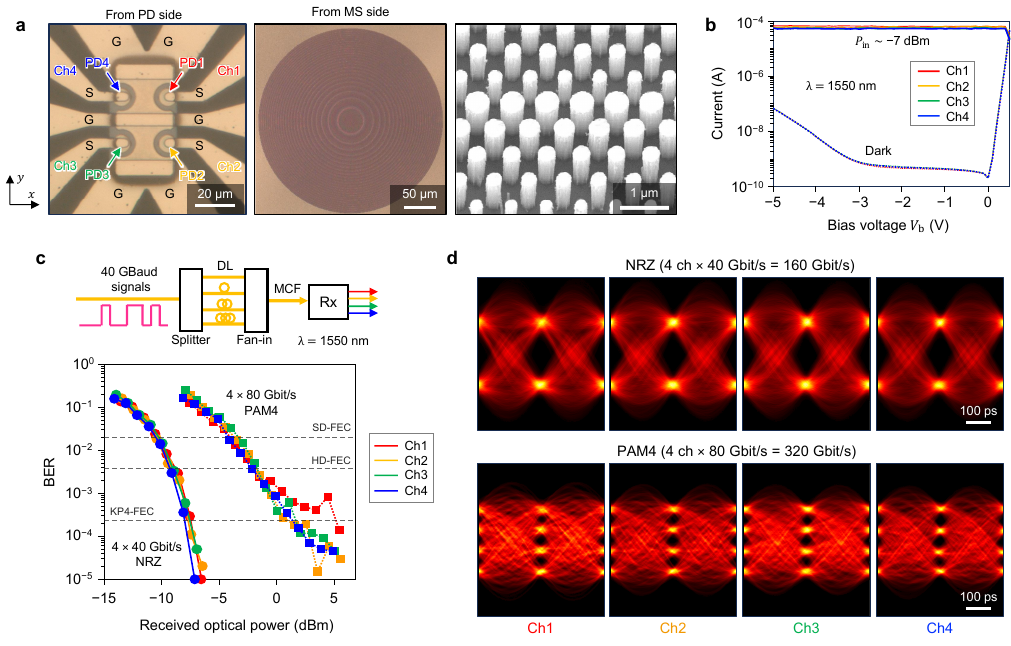}
\caption{
\textbf{Results of an ML-integrated 4-channel IM-DD receiver with an MCF.} 
\figlabel{a} Optical microscope and SEM images of the fabricated device, observed from the PDA and MS sides of the chip. 
The diameter of each PD is $D_\mathrm{PD}=8$~\textmu m. 
\figlabel{b} Measured I-V curves of PDs when 1550-nm wavelength light with $P_\mathrm{in}\sim-7$~dBm is incident from each core of an MCF. Dark currents of all PDs are also plotted.
\figlabel{c} Measured BERs of 40-Gbaud signals for all channels at 1550-nm wavelength as a function of the received optical power. The top panel shows a schematic of the setup. A 40-Gbaud signal is split into four paths with different-length DLs, transmitted through an MCF, and input to our receiver. 
\figlabel{d} Eye diagrams of received 4~Ch $\times$ 40-Gbit/s NRZ (160~Gbit/s) and 4~Ch $\times$ 80-Gbit/s PAM4 (320~Gbit/s) signals.
}
\label{fig:ML-PDA}
\end{figure}

\noindent
Figure~\ref{fig:ML-PDA} presents the results of extending our single-channel IM-DD receiver to four channels without additional components by exploiting the 2D spatial scalability of the surface-normal configuration. The fabricated device is shown in Fig.~\ref{fig:ML-PDA}a.
Here, four independent signals emitted from a four-core MCF are input to the device. The core pitch and MFD of the MCF are 40~{\textmu}m and 10.4~{\textmu}m, respectively. The separation between the MCF and MS is set to 631~{\textmu}m, so the MFD of each channel incident on the MS is around 120~{\textmu}m and overlaps with other channels. By designing the MS to operate as an ML with a focal length of 222~{\textmu}m, the near-field image at the MCF facet is demagnified by $\sim$0.54 and projected onto the PD plane. We can, therefore, detect four signals simultaneously using four-channel PDs with 8-{\textmu}m circular mesas placed at 22-{\textmu}m pitches (Fig.~\ref{fig:ML-PDA}a). 

Figure~\ref{fig:ML-PDA}b shows the measured I-V curves with and without light irradiation at a wavelength of 1550~nm. We confirm that identical characteristics are obtained for all PDs with crosstalk from other channels suppressed below $-20$~dB. 
Figures~\ref{fig:ML-PDA}c,d show the results of simultaneously receiving four-channel signals transmitted through the MCF.
Note that four signals are uncorrelated using optical delay lines (DLs) of different lengths, as shown in the top panel of Fig.~\ref{fig:ML-PDA}c.
BERs below the KP4-FEC threshold and clear eye diagrams are obtained for all four channels with 40-Gbit/s NRZ and 80-Gbit/s PAM4 formats, corresponding to 160-Gbit/s and 320-Gbit/s total data rates, respectively.

\subsection*{SVR with integrated polarization-sorting MS}

\begin{figure}[tb]
\centering\includegraphics{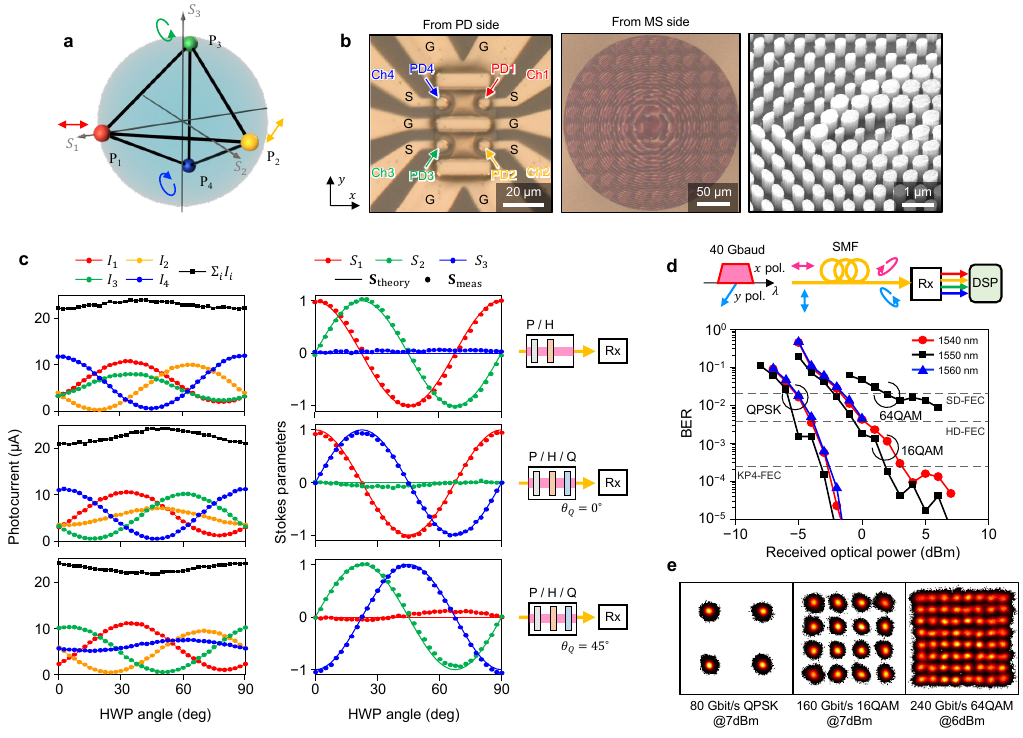}
\caption{
\textbf{SVR with an integrated polarization-sorting MS.} 
\figlabel{a} The four polarization bases $\mathrm{P_1, P_2, P_3, P_4}$, used to project the input signal. They are set to constitute a regular tetrahedron on the Poincaré sphere so that the receiver sensitivity is maximized. 
\figlabel{b} Optical microscope and SEM images of the fabricated device, observed from the PDA and MS sides of the chip. The diameter of each PD is $D_\mathrm{PD}=8$~\textmu m. 
\figlabel{c} Measured photocurrents $\mathbf{I}=(I_1, I_2, I_3, I_4)^t$ (left panel) and retrieved Stokes parameters $\mathbf{S}_\mathrm{meas}=(S_1, S_2, S_3)^t$ (center panel) as a function of the angle of an HWP, used to rotate the input polarization state in different configurations (right panel). Theoretical curves are plotted by the solid lines. 
P: polarizer. H: half-wave plate. Q: quarter-wave plate. 
\figlabel{d} Measured BERs of 40-Gbaud self-coherent signals as a function of the received optical power. Results at three different wavelengths of 1540, 1550, and 1560~nm are plotted for QPSK, 16QAM, and 64QAM formats. The top panel shows a schematic of the setup. A 40-Gbaud self-coherent signal is transmitted through an SMF and detected with a random polarization state by our SVR.
\figlabel{e} Constellation diagrams of received 80-Gbit/s QPSK, 160-Gbit/s 16QAM, and 240-Gbit/s 64QAM signals at 1550-nm wavelength.
}
\label{fig:SVR}
\end{figure}

\noindent
Figure~\ref{fig:SVR} shows the results of characterizing an SVR, which is capable of retrieving the polarization state of input light at high speed.
To enable polarization sorting functionality by an MS, we employed elliptical Si nanoposts (Fig.~\ref{fig:platform}a, right inset), arranged on a triangular lattice with a lattice constant of 700~nm.
As the light emitted from an SMF transmits through the MS, it is decomposed into four different polarization bases, $\mathrm{P}_1$, $\mathrm{P}_2$, $\mathrm{P}_3$, and $\mathrm{P}_4$, and focused to respective PDs (Fig.~\ref{fig:platform}b(iii)). 
From the four photocurrent signals, we can derive the full Stokes parameters of the input light through simple digital signal processing (DSP) \cite{Soma2024-fc}. 

It is proved that the maximum receiver sensitivity is obtained when the four polarization bases form a regular tetrahedron inscribed in the Poincaré sphere as shown in Fig.~\ref{fig:SVR}a \cite{Tyo2006-bi, Rubin2019-tb, Tanemura2020-nm, Soma2024-fc}. 
We therefore optimized the shape and rotation angle of elliptical Si nanoposts to achieve the Jones matrix distribution $\tilde{\mathbf{J}}(x, y)$ that provides such functionality \cite{Soma2024-fc} (see Methods and Extended Data Fig.~\ref{fig:SVR-design} for detailed design methods). Figure~\ref{fig:SVR}b shows the fabricated device. 

Figure~\ref{fig:SVR}c shows the measured photocurrent at four PDs, $\mathbf{I}=(I_1, I_2, I_3, I_4)^t$ (left panel), and the Stokes vector, $\mathbf{S}_\mathrm{meas}=(S_1, S_2, S_3)^t$ (center panel), which is retrieved from $\mathbf{I}$. 
The results are plotted as we rotate a half-wave plate (HWP) in three different configurations (right panel), so that the polarization state of input light is changed over the entire three-dimensional (3D) Stokes space.
The actual Stokes vector, $\mathbf{S}_\mathrm{theory}$, is also plotted with solid lines as a reference.
The average error $\langle|\mathbf{S}_\mathrm{meas}-\mathbf{S}_\mathrm{theory}|\rangle$ is as small as 0.071.

Since our SVR operates at high speed, it can be employed to demonstrate self-coherent signal transmission by sending a high-speed coherent signal on one polarization and non-modulated continuous-wave (CW) tone on the orthogonal polarization (see Methods and Extended Data Fig.~\ref{fig:setup}d for details). 
Although the polarization state evolves randomly during transmission through a non-polarization-maintaining SMF, such a polarization change can be removed through DSP by detecting the full Stokes vector using our SVR (Fig.~\ref{fig:SVR}d, top panel).

Figure~\ref{fig:SVR}d shows the measured BERs of 40-Gbaud self-coherent signals in quadrature phase shift keying (QPSK), 16-ary quadrature amplitude modulation (16QAM), and 64QAM formats, corresponding to 80-Gbit/s, 160-Gbit/s, and 240-Gbit/s data rates, respectively.
The constellation diagrams of the demodulated IQ signals at 1550-nm wavelength are shown in Fig.~\ref{fig:SVR}e.
Successful demodulation of up to 240-Gbit/s 64QAM signal at 1550-nm wavelength is achieved with a BER of less than $8.8\times 10^{-3}$, which is below the 20\% SD-FEC threshold.
Furthermore, a polarization-drift-resilient operation is experimentally confirmed by receiving 160-Gbit/s 16QAM signals for various input states of polarization (Extended Data Fig.~\ref{fig:SVR-pol}).

\subsection*{CR with an integrated MS}
\begin{figure}[tb]
\centering\includegraphics{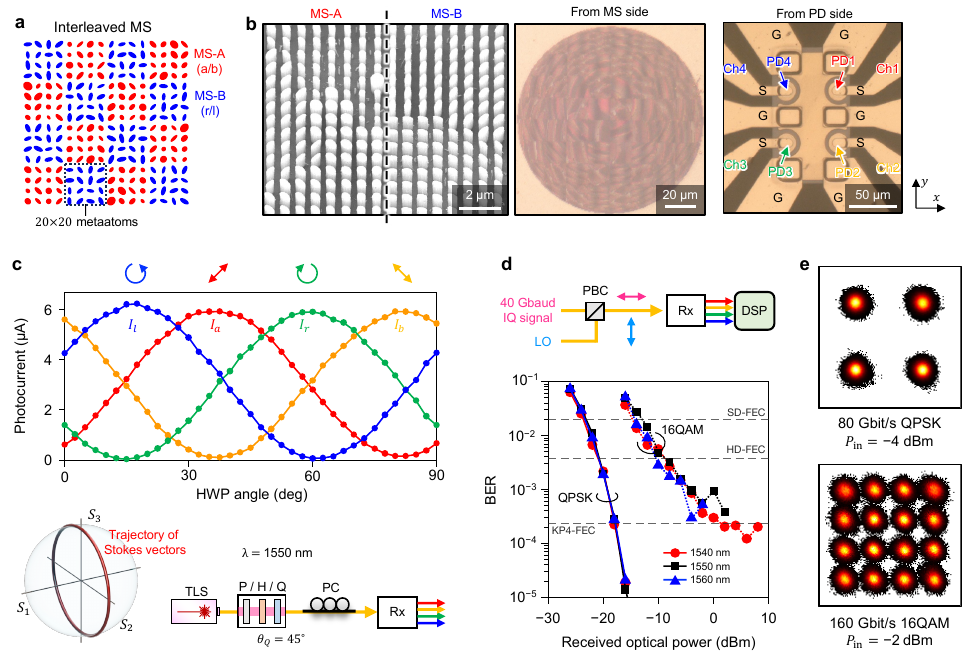}
\caption{
\textbf{CR with an integrated MS operating as an optical hybrid.} 
\figlabel{a} Schematic of the MS, where two independently designed sections, MS-A and MS-B, are interleaved. MS-A and MS-B split input light at $\pm 45^\circ$ linear polarization basis ($a$/$b$) and right- and left-handed circular polarization basis ($r$/$l$), respectively.
\figlabel{b} Optical microscope and SEM images of the fabricated device, observed from the MS and PDA sides. The diameter of each PD is $D_\mathrm{PD}=20$ \textmu m. 
\figlabel{c} Measured photocurrents $\mathbf{I}=(I_a, I_b, I_r, I_l)^t$ as a function of the HWP angle. The bottom panel shows the measurement setup and the trajectory of the input polarization state as we rotate the HWP.
PC: polarization controller.
\figlabel{d} Measured BERs of 80-Gbit/s QPSK and 160-Gbit/s 16QAM signals as a function of the received optical power at wavelengths of 1540, 1550, and 1560~nm.
The top panel shows the setup. 
\figlabel{e} Constellation diagrams of received 80-Gbit/s QPSK and 160-Gbit/s 16QAM signals at 1550-nm wavelength.
}
\label{fig:CR}
\end{figure}
\noindent
Finally, Fig.~\ref{fig:CR} shows the results of a CR with an integrated MS that functions as an optical hybrid.
As shown in Fig.~\ref{fig:CR}a, the MS in this device is constructed by interleaving two sections: MS-A and MS-B. 
Each section is comprised of $20\times20$ meta-atoms, arranged on a square lattice with a lattice constant of 700 nm.  
MS-A and MS-B are designed to split the input light at $\pm 45^\circ$ linear polarization bases ($a$/$b$) and the right- and left-handed circular polarization bases ($r$/$l$), respectively, and focus them to four separate PDs.

Here, a high-speed coherent signal in one polarization state is combined with CW local oscillator (LO) light in the orthogonal polarization state and input to the device (Fig.~\ref{fig:platform}b(iv)).
They are then split into four polarization components, $a$, $b$, $r$, and $l$, by transmitting through the MS and detected by the four PDs with relative phases of 0$^\circ$, 180$^\circ$, 90$^\circ$, and 270$^\circ$ between the signal and LO lightwaves, respectively.
Based on a similar principle as the polarizer-based configuration \cite{Soma2022-hl}, these PDs act as two sets of balanced PDs; by taking the differential signals between the $a$/$b$ and $r$/$l$ PDs (i.e., $I_a-I_b$ and $I_r-I_l$), we can retrieve the in-phase and quadrature components of input signal light, respectively. 
We should note that unlike the configuration using arrayed polarizers \cite{Soma2022-hl}, this scheme with a polarization-splitting MS does not suffer from a 3-dB intrinsic loss.
While we employ an off-chip polarization beam combiner (PBC) to combine the signal and LO in this work for the convenience of measurement, the MS could include the beam-combining functionality as well \cite{Komatsu2024-OFC} to realize a single-chip device.
The dimensions and rotation angles of the elliptical $\mathrm{\alpha}$-Si nanoposts in MS-A and MS-B were designed to realize desired focusing functionalities for corresponding polarization bases \cite{Arbabi2015-rv, Balthasar_Mueller2017-ex}.
Figure~\ref{fig:CR}b shows the fabricated device that contains four membrane PDs separated by 53~{\textmu}m, integrated with a 120-{\textmu}m-diameter MS on the other side. 

Figure~\ref{fig:CR}c shows the results of characterizing the polarization-splitting functionality of the MS. 
The photocurrents at four PDs are plotted as functions of the HWP angle (see the bottom panel in Fig.~\ref{fig:CR}c for the setup and trajectory of the input polarization state). Sinusoidal responses are obtained in agreement with the theory, demonstrating that both MS-A and MS-B sections operate properly. The extinction ratio of over 15~dB is obtained for all ports.
Finally, Figs.~\ref{fig:CR}d,e show the results of high-speed coherent detection experiments using our device.
Successful demodulation with a BER well below the 7\% hard-decision FEC (HD-FEC) threshold of $3.8\times 10^{-3}$ is achieved for both 80-Gbit/s QPSK and 160-Gbit/s 16QAM signals at 1540-, 1550-, and 1560-mm wavelengths (Fig.~\ref{fig:CR}d). Clear constellation diagrams of demodulated IQ signals are obtained for all cases (Fig.~\ref{fig:CR}e).

\section*{Conclusion}
\noindent
We have demonstrated a novel optical receiver platform with high spatial parallelism, consisting of an ultrathin dielectric MS and high-speed PDA integrated on a compact chip. Unlike conventional PIC-based receivers, our surface-normal platform offers a highly scalable solution for detecting a massive number of coherent optical channels without additional components. This was achieved through wafer bonding of an ultrafast ($>70$~GHz) membrane InGaAs/InP p-i-n PD layer to a SiO$_2$ substrate, along with backside integration of a functional MS composed of Si nanoposts.
The MS performs all the essential operations, including beam splitting, polarization sorting, and focusing, on normally incident light. Using the fabricated chip, we have successfully demonstrated simultaneous detection of 320-Gbit/s four-channel PAM4 signals by directly attaching a four-core MCF to the chip without using a fan-out device. Additionally, we achieved demodulation of high-speed coherent signals in various formats, including 80-Gbit/s QPSK, 160-Gbit/s 16QAM, and 240-Gbit/s 64QAM.

While the integration of MSs with active optoelectronic devices has been demonstrated for low-speed applications, such as ML-integrated image sensors \cite{Yun2021-mm, Uenoyama2022-fx, yang2023integrated}, 
mid-infrared photodiodes \cite{Wenger2022-en, Lien2024-cd}, and vertical-cavity surface-emitting lasers \cite{Xie2020-ga, Wen2021-uk, fu2023metasurface, zheng2025chip}, 
this work is, to our knowledge, the first to show that the integration of a functional MS with an ultrafast PDA unlocks new possibilities of MS technologies for advanced optical communication transceivers by providing 2D spatial scalability within a compact chip.
Moreover, our versatile platform holds great promise for a wide range of high-speed applications that leverage spatial parallelism of light, including ultra-dense optical interconnects \cite{Winzer2017-tx, Miller2017-gy, Winzer2023-qf}, free-space optical communication \cite{Wang2012-py, He2020-wc, Chen2021-sa, Horst2023-dp}, large-scale optical neural networks \cite{Hamerly2019-vs, Wang2022-qx, Chen2023-lk}, and coherent 3D imaging \cite{Martin2018-jx, Riemensberger2020-vi, Rogers2021-gk}.

\bibliography{references}

\begin{thebibliography}{83}%
\makeatletter
\providecommand \@ifxundefined [1]{%
 \@ifx{#1\undefined}
}%
\providecommand \@ifnum [1]{%
 \ifnum #1\expandafter \@firstoftwo
 \else \expandafter \@secondoftwo
 \fi
}%
\providecommand \@ifx [1]{%
 \ifx #1\expandafter \@firstoftwo
 \else \expandafter \@secondoftwo
 \fi
}%
\providecommand \natexlab [1]{#1}%
\providecommand \enquote  [1]{``#1''}%
\providecommand \bibnamefont  [1]{#1}%
\providecommand \bibfnamefont [1]{#1}%
\providecommand \citenamefont [1]{#1}%
\providecommand \href@noop [0]{\@secondoftwo}%
\providecommand \href [0]{\begingroup \@sanitize@url \@href}%
\providecommand \@href[1]{\@@startlink{#1}\@@href}%
\providecommand \@@href[1]{\endgroup#1\@@endlink}%
\providecommand \@sanitize@url [0]{\catcode `\\12\catcode `\$12\catcode `\&12\catcode `\#12\catcode `\^12\catcode `\_12\catcode `\%12\relax}%
\providecommand \@@startlink[1]{}%
\providecommand \@@endlink[0]{}%
\providecommand \url  [0]{\begingroup\@sanitize@url \@url }%
\providecommand \@url [1]{\endgroup\@href {#1}{\urlprefix }}%
\providecommand \urlprefix  [0]{URL }%
\providecommand \Eprint [0]{\href }%
\providecommand \doibase [0]{https://doi.org/}%
\providecommand \selectlanguage [0]{\@gobble}%
\providecommand \bibinfo  [0]{\@secondoftwo}%
\providecommand \bibfield  [0]{\@secondoftwo}%
\providecommand \translation [1]{[#1]}%
\providecommand \BibitemOpen [0]{}%
\providecommand \bibitemStop [0]{}%
\providecommand \bibitemNoStop [0]{.\EOS\space}%
\providecommand \EOS [0]{\spacefactor3000\relax}%
\providecommand \BibitemShut  [1]{\csname bibitem#1\endcsname}%
\let\auto@bib@innerbib\@empty
\bibitem [{\citenamefont {Yu}\ \emph {et~al.}(2011)\citenamefont {Yu}, \citenamefont {Genevet}, \citenamefont {Kats}, \citenamefont {Aieta}, \citenamefont {Tetienne}, \citenamefont {Capasso},\ and\ \citenamefont {Gaburro}}]{Yu2011-ey}%
  \BibitemOpen
  \bibfield  {author} {\bibinfo {author} {\bibfnamefont {N.}~\bibnamefont {Yu}}, \bibinfo {author} {\bibfnamefont {P.}~\bibnamefont {Genevet}}, \bibinfo {author} {\bibfnamefont {M.~A.}\ \bibnamefont {Kats}}, \bibinfo {author} {\bibfnamefont {F.}~\bibnamefont {Aieta}}, \bibinfo {author} {\bibfnamefont {J.-P.}\ \bibnamefont {Tetienne}}, \bibinfo {author} {\bibfnamefont {F.}~\bibnamefont {Capasso}},\ and\ \bibinfo {author} {\bibfnamefont {Z.}~\bibnamefont {Gaburro}},\ }\bibfield  {title} {\bibinfo {title} {Light propagation with phase discontinuities: generalized laws of reflection and refraction},\ }\href {https://doi.org/10.1126/science.1210713} {\bibfield  {journal} {\bibinfo  {journal} {Science}\ }\textbf {\bibinfo {volume} {334}},\ \bibinfo {pages} {333} (\bibinfo {year} {2011})}\BibitemShut {NoStop}%
\bibitem [{\citenamefont {Khorasaninejad}\ and\ \citenamefont {Capasso}(2017)}]{Khorasaninejad2017-sb}%
  \BibitemOpen
  \bibfield  {author} {\bibinfo {author} {\bibfnamefont {M.}~\bibnamefont {Khorasaninejad}}\ and\ \bibinfo {author} {\bibfnamefont {F.}~\bibnamefont {Capasso}},\ }\bibfield  {title} {\bibinfo {title} {Metalenses: {Versatile} multifunctional photonic components},\ }\href {https://doi.org/10.1126/science.aam8100} {\bibfield  {journal} {\bibinfo  {journal} {Science}\ }\textbf {\bibinfo {volume} {358}},\ \bibinfo {pages} {eaam8100} (\bibinfo {year} {2017})}\BibitemShut {NoStop}%
\bibitem [{\citenamefont {Arbabi}\ and\ \citenamefont {Faraon}(2022)}]{Arbabi2022-jk}%
  \BibitemOpen
  \bibfield  {author} {\bibinfo {author} {\bibfnamefont {A.}~\bibnamefont {Arbabi}}\ and\ \bibinfo {author} {\bibfnamefont {A.}~\bibnamefont {Faraon}},\ }\bibfield  {title} {\bibinfo {title} {Advances in optical metalenses},\ }\href {https://doi.org/10.1038/s41566-022-01108-6} {\bibfield  {journal} {\bibinfo  {journal} {Nat. Photonics}\ }\textbf {\bibinfo {volume} {17}},\ \bibinfo {pages} {16} (\bibinfo {year} {2022})}\BibitemShut {NoStop}%
\bibitem [{\citenamefont {Gu}\ \emph {et~al.}(2022)\citenamefont {Gu}, \citenamefont {Kim}, \citenamefont {Rivero-Baleine},\ and\ \citenamefont {Hu}}]{Gu2022-ky}%
  \BibitemOpen
  \bibfield  {author} {\bibinfo {author} {\bibfnamefont {T.}~\bibnamefont {Gu}}, \bibinfo {author} {\bibfnamefont {H.~J.}\ \bibnamefont {Kim}}, \bibinfo {author} {\bibfnamefont {C.}~\bibnamefont {Rivero-Baleine}},\ and\ \bibinfo {author} {\bibfnamefont {J.}~\bibnamefont {Hu}},\ }\bibfield  {title} {\bibinfo {title} {Reconfigurable metasurfaces towards commercial success},\ }\href {https://doi.org/10.1038/s41566-022-01099-4} {\bibfield  {journal} {\bibinfo  {journal} {Nat. Photonics}\ }\textbf {\bibinfo {volume} {17}},\ \bibinfo {pages} {48} (\bibinfo {year} {2022})}\BibitemShut {NoStop}%
\bibitem [{\citenamefont {Kuznetsov}\ \emph {et~al.}(2024)\citenamefont {Kuznetsov}, \citenamefont {Brongersma}, \citenamefont {Yao}, \citenamefont {Chen}, \citenamefont {Levy}, \citenamefont {Tsai}, \citenamefont {Zheludev}, \citenamefont {Faraon}, \citenamefont {Arbabi}, \citenamefont {Yu}, \citenamefont {Chanda}, \citenamefont {Crozier}, \citenamefont {Kildishev}, \citenamefont {Wang}, \citenamefont {Yang}, \citenamefont {Valentine}, \citenamefont {Genevet}, \citenamefont {Fan}, \citenamefont {Miller}, \citenamefont {Majumdar}, \citenamefont {Fröch}, \citenamefont {Brady}, \citenamefont {Heide}, \citenamefont {Veeraraghavan}, \citenamefont {Engheta}, \citenamefont {Alù}, \citenamefont {Polman}, \citenamefont {Atwater}, \citenamefont {Thureja}, \citenamefont {Paniagua-Dominguez}, \citenamefont {Ha}, \citenamefont {Barreda}, \citenamefont {Schuller}, \citenamefont {Staude}, \citenamefont {Grinblat}, \citenamefont {Kivshar}, \citenamefont {Peana}, \citenamefont {Yelin}, \citenamefont {Senichev},
  \citenamefont {Shalaev}, \citenamefont {Saha}, \citenamefont {Boltasseva}, \citenamefont {Rho}, \citenamefont {Oh}, \citenamefont {Kim}, \citenamefont {Park}, \citenamefont {Devlin},\ and\ \citenamefont {Pala}}]{Kuznetsov2024-lv}%
  \BibitemOpen
  \bibfield  {author} {\bibinfo {author} {\bibfnamefont {A.~I.}\ \bibnamefont {Kuznetsov}}, \bibinfo {author} {\bibfnamefont {M.~L.}\ \bibnamefont {Brongersma}}, \bibinfo {author} {\bibfnamefont {J.}~\bibnamefont {Yao}}, \bibinfo {author} {\bibfnamefont {M.~K.}\ \bibnamefont {Chen}}, \bibinfo {author} {\bibfnamefont {U.}~\bibnamefont {Levy}}, \bibinfo {author} {\bibfnamefont {D.~P.}\ \bibnamefont {Tsai}}, \bibinfo {author} {\bibfnamefont {N.~I.}\ \bibnamefont {Zheludev}}, \bibinfo {author} {\bibfnamefont {A.}~\bibnamefont {Faraon}}, \bibinfo {author} {\bibfnamefont {A.}~\bibnamefont {Arbabi}}, \bibinfo {author} {\bibfnamefont {N.}~\bibnamefont {Yu}}, \bibinfo {author} {\bibfnamefont {D.}~\bibnamefont {Chanda}}, \bibinfo {author} {\bibfnamefont {K.~B.}\ \bibnamefont {Crozier}}, \bibinfo {author} {\bibfnamefont {A.~V.}\ \bibnamefont {Kildishev}}, \bibinfo {author} {\bibfnamefont {H.}~\bibnamefont {Wang}}, \bibinfo {author} {\bibfnamefont {J.~K.~W.}\ \bibnamefont {Yang}}, \bibinfo {author} {\bibfnamefont
  {J.~G.}\ \bibnamefont {Valentine}}, \bibinfo {author} {\bibfnamefont {P.}~\bibnamefont {Genevet}}, \bibinfo {author} {\bibfnamefont {J.~A.}\ \bibnamefont {Fan}}, \bibinfo {author} {\bibfnamefont {O.~D.}\ \bibnamefont {Miller}}, \bibinfo {author} {\bibfnamefont {A.}~\bibnamefont {Majumdar}}, \bibinfo {author} {\bibfnamefont {J.~E.}\ \bibnamefont {Fröch}}, \bibinfo {author} {\bibfnamefont {D.}~\bibnamefont {Brady}}, \bibinfo {author} {\bibfnamefont {F.}~\bibnamefont {Heide}}, \bibinfo {author} {\bibfnamefont {A.}~\bibnamefont {Veeraraghavan}}, \bibinfo {author} {\bibfnamefont {N.}~\bibnamefont {Engheta}}, \bibinfo {author} {\bibfnamefont {A.}~\bibnamefont {Alù}}, \bibinfo {author} {\bibfnamefont {A.}~\bibnamefont {Polman}}, \bibinfo {author} {\bibfnamefont {H.~A.}\ \bibnamefont {Atwater}}, \bibinfo {author} {\bibfnamefont {P.}~\bibnamefont {Thureja}}, \bibinfo {author} {\bibfnamefont {R.}~\bibnamefont {Paniagua-Dominguez}}, \bibinfo {author} {\bibfnamefont {S.~T.}\ \bibnamefont {Ha}}, \bibinfo {author}
  {\bibfnamefont {A.~I.}\ \bibnamefont {Barreda}}, \bibinfo {author} {\bibfnamefont {J.~A.}\ \bibnamefont {Schuller}}, \bibinfo {author} {\bibfnamefont {I.}~\bibnamefont {Staude}}, \bibinfo {author} {\bibfnamefont {G.}~\bibnamefont {Grinblat}}, \bibinfo {author} {\bibfnamefont {Y.}~\bibnamefont {Kivshar}}, \bibinfo {author} {\bibfnamefont {S.}~\bibnamefont {Peana}}, \bibinfo {author} {\bibfnamefont {S.~F.}\ \bibnamefont {Yelin}}, \bibinfo {author} {\bibfnamefont {A.}~\bibnamefont {Senichev}}, \bibinfo {author} {\bibfnamefont {V.~M.}\ \bibnamefont {Shalaev}}, \bibinfo {author} {\bibfnamefont {S.}~\bibnamefont {Saha}}, \bibinfo {author} {\bibfnamefont {A.}~\bibnamefont {Boltasseva}}, \bibinfo {author} {\bibfnamefont {J.}~\bibnamefont {Rho}}, \bibinfo {author} {\bibfnamefont {D.~K.}\ \bibnamefont {Oh}}, \bibinfo {author} {\bibfnamefont {J.}~\bibnamefont {Kim}}, \bibinfo {author} {\bibfnamefont {J.}~\bibnamefont {Park}}, \bibinfo {author} {\bibfnamefont {R.}~\bibnamefont {Devlin}},\ and\ \bibinfo {author}
  {\bibfnamefont {R.~A.}\ \bibnamefont {Pala}},\ }\bibfield  {title} {\bibinfo {title} {Roadmap for optical metasurfaces},\ }\href {https://doi.org/10.1021/acsphotonics.3c00457} {\bibfield  {journal} {\bibinfo  {journal} {ACS Photonics}\ }\textbf {\bibinfo {volume} {11}},\ \bibinfo {pages} {816} (\bibinfo {year} {2024})}\BibitemShut {NoStop}%
\bibitem [{\citenamefont {Ha}\ \emph {et~al.}(2024)\citenamefont {Ha}, \citenamefont {Li}, \citenamefont {Yang}, \citenamefont {Demir}, \citenamefont {Brongersma},\ and\ \citenamefont {Kuznetsov}}]{Ha2024-mf}%
  \BibitemOpen
  \bibfield  {author} {\bibinfo {author} {\bibfnamefont {S.~T.}\ \bibnamefont {Ha}}, \bibinfo {author} {\bibfnamefont {Q.}~\bibnamefont {Li}}, \bibinfo {author} {\bibfnamefont {J.~K.~W.}\ \bibnamefont {Yang}}, \bibinfo {author} {\bibfnamefont {H.~V.}\ \bibnamefont {Demir}}, \bibinfo {author} {\bibfnamefont {M.~L.}\ \bibnamefont {Brongersma}},\ and\ \bibinfo {author} {\bibfnamefont {A.~I.}\ \bibnamefont {Kuznetsov}},\ }\bibfield  {title} {\bibinfo {title} {Optoelectronic metadevices},\ }\href {https://doi.org/10.1126/science.adm7442} {\bibfield  {journal} {\bibinfo  {journal} {Science}\ }\textbf {\bibinfo {volume} {386}},\ \bibinfo {pages} {eadm7442} (\bibinfo {year} {2024})}\BibitemShut {NoStop}%
\bibitem [{\citenamefont {Chen}\ \emph {et~al.}(2018)\citenamefont {Chen}, \citenamefont {Zhu}, \citenamefont {Sanjeev}, \citenamefont {Khorasaninejad}, \citenamefont {Shi}, \citenamefont {Lee},\ and\ \citenamefont {Capasso}}]{Chen2018-lu}%
  \BibitemOpen
  \bibfield  {author} {\bibinfo {author} {\bibfnamefont {W.~T.}\ \bibnamefont {Chen}}, \bibinfo {author} {\bibfnamefont {A.~Y.}\ \bibnamefont {Zhu}}, \bibinfo {author} {\bibfnamefont {V.}~\bibnamefont {Sanjeev}}, \bibinfo {author} {\bibfnamefont {M.}~\bibnamefont {Khorasaninejad}}, \bibinfo {author} {\bibfnamefont {Z.}~\bibnamefont {Shi}}, \bibinfo {author} {\bibfnamefont {E.}~\bibnamefont {Lee}},\ and\ \bibinfo {author} {\bibfnamefont {F.}~\bibnamefont {Capasso}},\ }\bibfield  {title} {\bibinfo {title} {A broadband achromatic metalens for focusing and imaging in the visible},\ }\href {https://doi.org/10.1038/s41565-017-0034-6} {\bibfield  {journal} {\bibinfo  {journal} {Nat. Nanotechnol.}\ }\textbf {\bibinfo {volume} {13}},\ \bibinfo {pages} {220} (\bibinfo {year} {2018})}\BibitemShut {NoStop}%
\bibitem [{\citenamefont {Shrestha}\ \emph {et~al.}(2018)\citenamefont {Shrestha}, \citenamefont {Overvig}, \citenamefont {Lu}, \citenamefont {Stein},\ and\ \citenamefont {Yu}}]{Shrestha2018-cx}%
  \BibitemOpen
  \bibfield  {author} {\bibinfo {author} {\bibfnamefont {S.}~\bibnamefont {Shrestha}}, \bibinfo {author} {\bibfnamefont {A.~C.}\ \bibnamefont {Overvig}}, \bibinfo {author} {\bibfnamefont {M.}~\bibnamefont {Lu}}, \bibinfo {author} {\bibfnamefont {A.}~\bibnamefont {Stein}},\ and\ \bibinfo {author} {\bibfnamefont {N.}~\bibnamefont {Yu}},\ }\bibfield  {title} {\bibinfo {title} {Broadband achromatic dielectric metalenses},\ }\href {https://doi.org/10.1038/s41377-018-0078-x} {\bibfield  {journal} {\bibinfo  {journal} {Light Sci Appl}\ }\textbf {\bibinfo {volume} {7}},\ \bibinfo {pages} {85} (\bibinfo {year} {2018})}\BibitemShut {NoStop}%
\bibitem [{\citenamefont {Wang}\ \emph {et~al.}(2018)\citenamefont {Wang}, \citenamefont {Wu}, \citenamefont {Su}, \citenamefont {Lai}, \citenamefont {Chen}, \citenamefont {Kuo}, \citenamefont {Chen}, \citenamefont {Chen}, \citenamefont {Huang}, \citenamefont {Wang}, \citenamefont {Lin}, \citenamefont {Kuan}, \citenamefont {Li}, \citenamefont {Wang}, \citenamefont {Zhu},\ and\ \citenamefont {Tsai}}]{Wang2018-cm}%
  \BibitemOpen
  \bibfield  {author} {\bibinfo {author} {\bibfnamefont {S.}~\bibnamefont {Wang}}, \bibinfo {author} {\bibfnamefont {P.~C.}\ \bibnamefont {Wu}}, \bibinfo {author} {\bibfnamefont {V.-C.}\ \bibnamefont {Su}}, \bibinfo {author} {\bibfnamefont {Y.-C.}\ \bibnamefont {Lai}}, \bibinfo {author} {\bibfnamefont {M.-K.}\ \bibnamefont {Chen}}, \bibinfo {author} {\bibfnamefont {H.~Y.}\ \bibnamefont {Kuo}}, \bibinfo {author} {\bibfnamefont {B.~H.}\ \bibnamefont {Chen}}, \bibinfo {author} {\bibfnamefont {Y.~H.}\ \bibnamefont {Chen}}, \bibinfo {author} {\bibfnamefont {T.-T.}\ \bibnamefont {Huang}}, \bibinfo {author} {\bibfnamefont {J.-H.}\ \bibnamefont {Wang}}, \bibinfo {author} {\bibfnamefont {R.-M.}\ \bibnamefont {Lin}}, \bibinfo {author} {\bibfnamefont {C.-H.}\ \bibnamefont {Kuan}}, \bibinfo {author} {\bibfnamefont {T.}~\bibnamefont {Li}}, \bibinfo {author} {\bibfnamefont {Z.}~\bibnamefont {Wang}}, \bibinfo {author} {\bibfnamefont {S.}~\bibnamefont {Zhu}},\ and\ \bibinfo {author} {\bibfnamefont {D.~P.}\ \bibnamefont
  {Tsai}},\ }\bibfield  {title} {\bibinfo {title} {A broadband achromatic metalens in the visible},\ }\href {https://doi.org/10.1038/s41565-017-0052-4} {\bibfield  {journal} {\bibinfo  {journal} {Nat. Nanotechnol.}\ }\textbf {\bibinfo {volume} {13}},\ \bibinfo {pages} {227} (\bibinfo {year} {2018})}\BibitemShut {NoStop}%
\bibitem [{\citenamefont {Chen}\ \emph {et~al.}(2019)\citenamefont {Chen}, \citenamefont {Zhu}, \citenamefont {Sisler}, \citenamefont {Bharwani},\ and\ \citenamefont {Capasso}}]{Chen2019-eo}%
  \BibitemOpen
  \bibfield  {author} {\bibinfo {author} {\bibfnamefont {W.~T.}\ \bibnamefont {Chen}}, \bibinfo {author} {\bibfnamefont {A.~Y.}\ \bibnamefont {Zhu}}, \bibinfo {author} {\bibfnamefont {J.}~\bibnamefont {Sisler}}, \bibinfo {author} {\bibfnamefont {Z.}~\bibnamefont {Bharwani}},\ and\ \bibinfo {author} {\bibfnamefont {F.}~\bibnamefont {Capasso}},\ }\bibfield  {title} {\bibinfo {title} {A broadband achromatic polarization-insensitive metalens consisting of anisotropic nanostructures},\ }\href {https://doi.org/10.1038/s41467-019-08305-y} {\bibfield  {journal} {\bibinfo  {journal} {Nat. Commun.}\ }\textbf {\bibinfo {volume} {10}},\ \bibinfo {pages} {355} (\bibinfo {year} {2019})}\BibitemShut {NoStop}%
\bibitem [{\citenamefont {Hu}\ \emph {et~al.}(2023)\citenamefont {Hu}, \citenamefont {Jiang}, \citenamefont {Zhang}, \citenamefont {Yang}, \citenamefont {Ou}, \citenamefont {Li}, \citenamefont {Kong}, \citenamefont {Liu}, \citenamefont {Qiu},\ and\ \citenamefont {Duan}}]{Hu2023-ni}%
  \BibitemOpen
  \bibfield  {author} {\bibinfo {author} {\bibfnamefont {Y.}~\bibnamefont {Hu}}, \bibinfo {author} {\bibfnamefont {Y.}~\bibnamefont {Jiang}}, \bibinfo {author} {\bibfnamefont {Y.}~\bibnamefont {Zhang}}, \bibinfo {author} {\bibfnamefont {X.}~\bibnamefont {Yang}}, \bibinfo {author} {\bibfnamefont {X.}~\bibnamefont {Ou}}, \bibinfo {author} {\bibfnamefont {L.}~\bibnamefont {Li}}, \bibinfo {author} {\bibfnamefont {X.}~\bibnamefont {Kong}}, \bibinfo {author} {\bibfnamefont {X.}~\bibnamefont {Liu}}, \bibinfo {author} {\bibfnamefont {C.-W.}\ \bibnamefont {Qiu}},\ and\ \bibinfo {author} {\bibfnamefont {H.}~\bibnamefont {Duan}},\ }\bibfield  {title} {\bibinfo {title} {Asymptotic dispersion engineering for ultra-broadband meta-optics},\ }\href {https://doi.org/10.1038/s41467-023-42268-5} {\bibfield  {journal} {\bibinfo  {journal} {Nat. Commun.}\ }\textbf {\bibinfo {volume} {14}},\ \bibinfo {pages} {6649} (\bibinfo {year} {2023})}\BibitemShut {NoStop}%
\bibitem [{\citenamefont {Rubin}\ \emph {et~al.}(2019)\citenamefont {Rubin}, \citenamefont {D'Aversa}, \citenamefont {Chevalier}, \citenamefont {Shi}, \citenamefont {Chen},\ and\ \citenamefont {Capasso}}]{Rubin2019-tb}%
  \BibitemOpen
  \bibfield  {author} {\bibinfo {author} {\bibfnamefont {N.~A.}\ \bibnamefont {Rubin}}, \bibinfo {author} {\bibfnamefont {G.}~\bibnamefont {D'Aversa}}, \bibinfo {author} {\bibfnamefont {P.}~\bibnamefont {Chevalier}}, \bibinfo {author} {\bibfnamefont {Z.}~\bibnamefont {Shi}}, \bibinfo {author} {\bibfnamefont {W.~T.}\ \bibnamefont {Chen}},\ and\ \bibinfo {author} {\bibfnamefont {F.}~\bibnamefont {Capasso}},\ }\bibfield  {title} {\bibinfo {title} {Matrix {Fourier} optics enables a compact full-{Stokes} polarization camera},\ }\href {https://doi.org/10.1126/science.aax1839} {\bibfield  {journal} {\bibinfo  {journal} {Science}\ }\textbf {\bibinfo {volume} {365}},\ \bibinfo {pages} {eaax1839} (\bibinfo {year} {2019})}\BibitemShut {NoStop}%
\bibitem [{\citenamefont {Arbabi}\ \emph {et~al.}(2018)\citenamefont {Arbabi}, \citenamefont {Kamali}, \citenamefont {Arbabi},\ and\ \citenamefont {Faraon}}]{Arbabi2018-vz}%
  \BibitemOpen
  \bibfield  {author} {\bibinfo {author} {\bibfnamefont {E.}~\bibnamefont {Arbabi}}, \bibinfo {author} {\bibfnamefont {S.~M.}\ \bibnamefont {Kamali}}, \bibinfo {author} {\bibfnamefont {A.}~\bibnamefont {Arbabi}},\ and\ \bibinfo {author} {\bibfnamefont {A.}~\bibnamefont {Faraon}},\ }\bibfield  {title} {\bibinfo {title} {{Full-Stokes} imaging polarimetry using dielectric metasurfaces},\ }\href {https://doi.org/10.1021/acsphotonics.8b00362} {\bibfield  {journal} {\bibinfo  {journal} {ACS Photonics}\ }\textbf {\bibinfo {volume} {5}},\ \bibinfo {pages} {3132} (\bibinfo {year} {2018})}\BibitemShut {NoStop}%
\bibitem [{\citenamefont {Pors}\ \emph {et~al.}(2015)\citenamefont {Pors}, \citenamefont {Nielsen},\ and\ \citenamefont {Bozhevolnyi}}]{Pors2015-il}%
  \BibitemOpen
  \bibfield  {author} {\bibinfo {author} {\bibfnamefont {A.}~\bibnamefont {Pors}}, \bibinfo {author} {\bibfnamefont {M.~G.}\ \bibnamefont {Nielsen}},\ and\ \bibinfo {author} {\bibfnamefont {S.~I.}\ \bibnamefont {Bozhevolnyi}},\ }\bibfield  {title} {\bibinfo {title} {Plasmonic metagratings for simultaneous determination of {Stokes} parameters},\ }\href {https://doi.org/10.1364/OPTICA.2.000716} {\bibfield  {journal} {\bibinfo  {journal} {Optica}\ }\textbf {\bibinfo {volume} {2}},\ \bibinfo {pages} {716} (\bibinfo {year} {2015})}\BibitemShut {NoStop}%
\bibitem [{\citenamefont {Yang}\ \emph {et~al.}(2018)\citenamefont {Yang}, \citenamefont {Wang}, \citenamefont {Wang}, \citenamefont {Feng}, \citenamefont {Zhao}, \citenamefont {Wan}, \citenamefont {Zhu}, \citenamefont {Liu}, \citenamefont {Huang}, \citenamefont {Xia},\ and\ \citenamefont {Wegener}}]{Yang2018-yh}%
  \BibitemOpen
  \bibfield  {author} {\bibinfo {author} {\bibfnamefont {Z.}~\bibnamefont {Yang}}, \bibinfo {author} {\bibfnamefont {Z.}~\bibnamefont {Wang}}, \bibinfo {author} {\bibfnamefont {Y.}~\bibnamefont {Wang}}, \bibinfo {author} {\bibfnamefont {X.}~\bibnamefont {Feng}}, \bibinfo {author} {\bibfnamefont {M.}~\bibnamefont {Zhao}}, \bibinfo {author} {\bibfnamefont {Z.}~\bibnamefont {Wan}}, \bibinfo {author} {\bibfnamefont {L.}~\bibnamefont {Zhu}}, \bibinfo {author} {\bibfnamefont {J.}~\bibnamefont {Liu}}, \bibinfo {author} {\bibfnamefont {Y.}~\bibnamefont {Huang}}, \bibinfo {author} {\bibfnamefont {J.}~\bibnamefont {Xia}},\ and\ \bibinfo {author} {\bibfnamefont {M.}~\bibnamefont {Wegener}},\ }\bibfield  {title} {\bibinfo {title} {Generalized {Hartmann-Shack} array of dielectric metalens sub-arrays for polarimetric beam profiling},\ }\href {https://doi.org/10.1038/s41467-018-07056-6} {\bibfield  {journal} {\bibinfo  {journal} {Nat. Commun.}\ }\textbf {\bibinfo {volume} {9}},\ \bibinfo {pages} {4607} (\bibinfo {year}
  {2018})}\BibitemShut {NoStop}%
\bibitem [{\citenamefont {Shah}\ \emph {et~al.}(2022)\citenamefont {Shah}, \citenamefont {Dada}, \citenamefont {Grant}, \citenamefont {Cumming}, \citenamefont {Altuzarra}, \citenamefont {Nowack}, \citenamefont {Lyons}, \citenamefont {Clerici},\ and\ \citenamefont {Faccio}}]{Shah2022-fm}%
  \BibitemOpen
  \bibfield  {author} {\bibinfo {author} {\bibfnamefont {Y.~D.}\ \bibnamefont {Shah}}, \bibinfo {author} {\bibfnamefont {A.~C.}\ \bibnamefont {Dada}}, \bibinfo {author} {\bibfnamefont {J.~P.}\ \bibnamefont {Grant}}, \bibinfo {author} {\bibfnamefont {D.~R.~S.}\ \bibnamefont {Cumming}}, \bibinfo {author} {\bibfnamefont {C.}~\bibnamefont {Altuzarra}}, \bibinfo {author} {\bibfnamefont {T.~S.}\ \bibnamefont {Nowack}}, \bibinfo {author} {\bibfnamefont {A.}~\bibnamefont {Lyons}}, \bibinfo {author} {\bibfnamefont {M.}~\bibnamefont {Clerici}},\ and\ \bibinfo {author} {\bibfnamefont {D.}~\bibnamefont {Faccio}},\ }\bibfield  {title} {\bibinfo {title} {An all-dielectric metasurface polarimeter},\ }\href {https://doi.org/10.1021/acsphotonics.2c00395} {\bibfield  {journal} {\bibinfo  {journal} {ACS Photonics}\ }\textbf {\bibinfo {volume} {9}},\ \bibinfo {pages} {3245} (\bibinfo {year} {2022})}\BibitemShut {NoStop}%
\bibitem [{\citenamefont {Soma}\ \emph {et~al.}(2024)\citenamefont {Soma}, \citenamefont {Komatsu}, \citenamefont {Ren}, \citenamefont {Nakano},\ and\ \citenamefont {Tanemura}}]{Soma2024-fc}%
  \BibitemOpen
  \bibfield  {author} {\bibinfo {author} {\bibfnamefont {G.}~\bibnamefont {Soma}}, \bibinfo {author} {\bibfnamefont {K.}~\bibnamefont {Komatsu}}, \bibinfo {author} {\bibfnamefont {C.}~\bibnamefont {Ren}}, \bibinfo {author} {\bibfnamefont {Y.}~\bibnamefont {Nakano}},\ and\ \bibinfo {author} {\bibfnamefont {T.}~\bibnamefont {Tanemura}},\ }\bibfield  {title} {\bibinfo {title} {Metasurface-enabled non-orthogonal four-output polarization splitter for non-redundant full-stokes imaging},\ }\href {https://doi.org/10.1364/oe.529389} {\bibfield  {journal} {\bibinfo  {journal} {Opt. Express}\ }\textbf {\bibinfo {volume} {32}},\ \bibinfo {pages} {34207} (\bibinfo {year} {2024})}\BibitemShut {NoStop}%
\bibitem [{\citenamefont {Faraji-Dana}\ \emph {et~al.}(2018)\citenamefont {Faraji-Dana}, \citenamefont {Arbabi}, \citenamefont {Arbabi}, \citenamefont {Kamali}, \citenamefont {Kwon},\ and\ \citenamefont {Faraon}}]{Faraji-Dana2018-xc}%
  \BibitemOpen
  \bibfield  {author} {\bibinfo {author} {\bibfnamefont {M.}~\bibnamefont {Faraji-Dana}}, \bibinfo {author} {\bibfnamefont {E.}~\bibnamefont {Arbabi}}, \bibinfo {author} {\bibfnamefont {A.}~\bibnamefont {Arbabi}}, \bibinfo {author} {\bibfnamefont {S.~M.}\ \bibnamefont {Kamali}}, \bibinfo {author} {\bibfnamefont {H.}~\bibnamefont {Kwon}},\ and\ \bibinfo {author} {\bibfnamefont {A.}~\bibnamefont {Faraon}},\ }\bibfield  {title} {\bibinfo {title} {Compact folded metasurface spectrometer},\ }\href {https://doi.org/10.1038/s41467-018-06495-5} {\bibfield  {journal} {\bibinfo  {journal} {Nat. Commun.}\ }\textbf {\bibinfo {volume} {9}},\ \bibinfo {pages} {4196} (\bibinfo {year} {2018})}\BibitemShut {NoStop}%
\bibitem [{\citenamefont {Zhu}\ \emph {et~al.}(2019)\citenamefont {Zhu}, \citenamefont {Chen}, \citenamefont {Sisler}, \citenamefont {Yousef}, \citenamefont {Lee}, \citenamefont {Huang}, \citenamefont {Qiu},\ and\ \citenamefont {Capasso}}]{Zhu2019-xe}%
  \BibitemOpen
  \bibfield  {author} {\bibinfo {author} {\bibfnamefont {A.~Y.}\ \bibnamefont {Zhu}}, \bibinfo {author} {\bibfnamefont {W.~T.}\ \bibnamefont {Chen}}, \bibinfo {author} {\bibfnamefont {J.}~\bibnamefont {Sisler}}, \bibinfo {author} {\bibfnamefont {K.~M.~A.}\ \bibnamefont {Yousef}}, \bibinfo {author} {\bibfnamefont {E.}~\bibnamefont {Lee}}, \bibinfo {author} {\bibfnamefont {Y.-W.}\ \bibnamefont {Huang}}, \bibinfo {author} {\bibfnamefont {C.-W.}\ \bibnamefont {Qiu}},\ and\ \bibinfo {author} {\bibfnamefont {F.}~\bibnamefont {Capasso}},\ }\bibfield  {title} {\bibinfo {title} {Compact aberration‐corrected spectrometers in the visible using dispersion‐tailored metasurfaces},\ }\href {https://doi.org/10.1002/adom.201801144} {\bibfield  {journal} {\bibinfo  {journal} {Adv. Opt. Mater.}\ }\textbf {\bibinfo {volume} {7}},\ \bibinfo {pages} {1801144} (\bibinfo {year} {2019})}\BibitemShut {NoStop}%
\bibitem [{\citenamefont {Miyata}\ \emph {et~al.}(2021)\citenamefont {Miyata}, \citenamefont {Nemoto}, \citenamefont {Shikama}, \citenamefont {Kobayashi},\ and\ \citenamefont {Hashimoto}}]{Miyata2021-cg}%
  \BibitemOpen
  \bibfield  {author} {\bibinfo {author} {\bibfnamefont {M.}~\bibnamefont {Miyata}}, \bibinfo {author} {\bibfnamefont {N.}~\bibnamefont {Nemoto}}, \bibinfo {author} {\bibfnamefont {K.}~\bibnamefont {Shikama}}, \bibinfo {author} {\bibfnamefont {F.}~\bibnamefont {Kobayashi}},\ and\ \bibinfo {author} {\bibfnamefont {T.}~\bibnamefont {Hashimoto}},\ }\bibfield  {title} {\bibinfo {title} {Full-color-sorting metalenses for high-sensitivity image sensors},\ }\href {https://doi.org/10.1364/OPTICA.444255} {\bibfield  {journal} {\bibinfo  {journal} {Optica}\ }\textbf {\bibinfo {volume} {8}},\ \bibinfo {pages} {1596} (\bibinfo {year} {2021})}\BibitemShut {NoStop}%
\bibitem [{\citenamefont {Zou}\ \emph {et~al.}(2022)\citenamefont {Zou}, \citenamefont {Zhang}, \citenamefont {Lin}, \citenamefont {Gong}, \citenamefont {Wang}, \citenamefont {Zhu},\ and\ \citenamefont {Wang}}]{Zou2022-xp}%
  \BibitemOpen
  \bibfield  {author} {\bibinfo {author} {\bibfnamefont {X.}~\bibnamefont {Zou}}, \bibinfo {author} {\bibfnamefont {Y.}~\bibnamefont {Zhang}}, \bibinfo {author} {\bibfnamefont {R.}~\bibnamefont {Lin}}, \bibinfo {author} {\bibfnamefont {G.}~\bibnamefont {Gong}}, \bibinfo {author} {\bibfnamefont {S.}~\bibnamefont {Wang}}, \bibinfo {author} {\bibfnamefont {S.}~\bibnamefont {Zhu}},\ and\ \bibinfo {author} {\bibfnamefont {Z.}~\bibnamefont {Wang}},\ }\bibfield  {title} {\bibinfo {title} {Pixel-level {Bayer-type} colour router based on metasurfaces},\ }\href {https://doi.org/10.1038/s41467-022-31019-7} {\bibfield  {journal} {\bibinfo  {journal} {Nat. Commun.}\ }\textbf {\bibinfo {volume} {13}},\ \bibinfo {pages} {3288} (\bibinfo {year} {2022})}\BibitemShut {NoStop}%
\bibitem [{\citenamefont {Ni}\ \emph {et~al.}(2013)\citenamefont {Ni}, \citenamefont {Kildishev},\ and\ \citenamefont {Shalaev}}]{Ni2013-be}%
  \BibitemOpen
  \bibfield  {author} {\bibinfo {author} {\bibfnamefont {X.}~\bibnamefont {Ni}}, \bibinfo {author} {\bibfnamefont {A.~V.}\ \bibnamefont {Kildishev}},\ and\ \bibinfo {author} {\bibfnamefont {V.~M.}\ \bibnamefont {Shalaev}},\ }\bibfield  {title} {\bibinfo {title} {Metasurface holograms for visible light},\ }\href {https://doi.org/10.1038/ncomms3807} {\bibfield  {journal} {\bibinfo  {journal} {Nat. Commun.}\ }\textbf {\bibinfo {volume} {4}},\ \bibinfo {pages} {2807} (\bibinfo {year} {2013})}\BibitemShut {NoStop}%
\bibitem [{\citenamefont {Zheng}\ \emph {et~al.}(2015)\citenamefont {Zheng}, \citenamefont {M{\"u}hlenbernd}, \citenamefont {Kenney}, \citenamefont {Li}, \citenamefont {Zentgraf},\ and\ \citenamefont {Zhang}}]{Zheng2015-xm}%
  \BibitemOpen
  \bibfield  {author} {\bibinfo {author} {\bibfnamefont {G.}~\bibnamefont {Zheng}}, \bibinfo {author} {\bibfnamefont {H.}~\bibnamefont {M{\"u}hlenbernd}}, \bibinfo {author} {\bibfnamefont {M.}~\bibnamefont {Kenney}}, \bibinfo {author} {\bibfnamefont {G.}~\bibnamefont {Li}}, \bibinfo {author} {\bibfnamefont {T.}~\bibnamefont {Zentgraf}},\ and\ \bibinfo {author} {\bibfnamefont {S.}~\bibnamefont {Zhang}},\ }\bibfield  {title} {\bibinfo {title} {Metasurface holograms reaching 80\% efficiency},\ }\href {https://doi.org/10.1038/nnano.2015.2} {\bibfield  {journal} {\bibinfo  {journal} {Nat. Nanotechnol.}\ }\textbf {\bibinfo {volume} {10}},\ \bibinfo {pages} {308} (\bibinfo {year} {2015})}\BibitemShut {NoStop}%
\bibitem [{\citenamefont {Wang}\ \emph {et~al.}(2016)\citenamefont {Wang}, \citenamefont {Dong}, \citenamefont {Li}, \citenamefont {Yang}, \citenamefont {Sun}, \citenamefont {Chen}, \citenamefont {Song}, \citenamefont {Xu}, \citenamefont {Chu}, \citenamefont {Xiao}, \citenamefont {Gong},\ and\ \citenamefont {Li}}]{Wang2016-aj}%
  \BibitemOpen
  \bibfield  {author} {\bibinfo {author} {\bibfnamefont {B.}~\bibnamefont {Wang}}, \bibinfo {author} {\bibfnamefont {F.}~\bibnamefont {Dong}}, \bibinfo {author} {\bibfnamefont {Q.-T.}\ \bibnamefont {Li}}, \bibinfo {author} {\bibfnamefont {D.}~\bibnamefont {Yang}}, \bibinfo {author} {\bibfnamefont {C.}~\bibnamefont {Sun}}, \bibinfo {author} {\bibfnamefont {J.}~\bibnamefont {Chen}}, \bibinfo {author} {\bibfnamefont {Z.}~\bibnamefont {Song}}, \bibinfo {author} {\bibfnamefont {L.}~\bibnamefont {Xu}}, \bibinfo {author} {\bibfnamefont {W.}~\bibnamefont {Chu}}, \bibinfo {author} {\bibfnamefont {Y.-F.}\ \bibnamefont {Xiao}}, \bibinfo {author} {\bibfnamefont {Q.}~\bibnamefont {Gong}},\ and\ \bibinfo {author} {\bibfnamefont {Y.}~\bibnamefont {Li}},\ }\bibfield  {title} {\bibinfo {title} {Visible-frequency dielectric metasurfaces for multiwavelength achromatic and highly dispersive holograms},\ }\href {https://doi.org/10.1021/acs.nanolett.6b02326} {\bibfield  {journal} {\bibinfo  {journal} {Nano Lett.}\ }\textbf
  {\bibinfo {volume} {16}},\ \bibinfo {pages} {5235} (\bibinfo {year} {2016})}\BibitemShut {NoStop}%
\bibitem [{\citenamefont {Arbabi}\ \emph {et~al.}(2015)\citenamefont {Arbabi}, \citenamefont {Horie}, \citenamefont {Bagheri},\ and\ \citenamefont {Faraon}}]{Arbabi2015-rv}%
  \BibitemOpen
  \bibfield  {author} {\bibinfo {author} {\bibfnamefont {A.}~\bibnamefont {Arbabi}}, \bibinfo {author} {\bibfnamefont {Y.}~\bibnamefont {Horie}}, \bibinfo {author} {\bibfnamefont {M.}~\bibnamefont {Bagheri}},\ and\ \bibinfo {author} {\bibfnamefont {A.}~\bibnamefont {Faraon}},\ }\bibfield  {title} {\bibinfo {title} {Dielectric metasurfaces for complete control of phase and polarization with subwavelength spatial resolution and high transmission},\ }\href {https://doi.org/10.1038/nnano.2015.186} {\bibfield  {journal} {\bibinfo  {journal} {Nat. Nanotechnol.}\ }\textbf {\bibinfo {volume} {10}},\ \bibinfo {pages} {937} (\bibinfo {year} {2015})}\BibitemShut {NoStop}%
\bibitem [{\citenamefont {Balthasar~Mueller}\ \emph {et~al.}(2017)\citenamefont {Balthasar~Mueller}, \citenamefont {Rubin}, \citenamefont {Devlin}, \citenamefont {Groever},\ and\ \citenamefont {Capasso}}]{Balthasar_Mueller2017-ex}%
  \BibitemOpen
  \bibfield  {author} {\bibinfo {author} {\bibfnamefont {J.~P.}\ \bibnamefont {Balthasar~Mueller}}, \bibinfo {author} {\bibfnamefont {N.~A.}\ \bibnamefont {Rubin}}, \bibinfo {author} {\bibfnamefont {R.~C.}\ \bibnamefont {Devlin}}, \bibinfo {author} {\bibfnamefont {B.}~\bibnamefont {Groever}},\ and\ \bibinfo {author} {\bibfnamefont {F.}~\bibnamefont {Capasso}},\ }\bibfield  {title} {\bibinfo {title} {Metasurface polarization optics: independent phase control of arbitrary orthogonal states of polarization},\ }\href {https://doi.org/10.1103/PhysRevLett.118.113901} {\bibfield  {journal} {\bibinfo  {journal} {Phys. Rev. Lett.}\ }\textbf {\bibinfo {volume} {118}},\ \bibinfo {pages} {113901} (\bibinfo {year} {2017})}\BibitemShut {NoStop}%
\bibitem [{\citenamefont {Ren}\ \emph {et~al.}(2020)\citenamefont {Ren}, \citenamefont {Fang}, \citenamefont {Jang}, \citenamefont {B{\"u}rger}, \citenamefont {Rho},\ and\ \citenamefont {Maier}}]{Ren2020-bs}%
  \BibitemOpen
  \bibfield  {author} {\bibinfo {author} {\bibfnamefont {H.}~\bibnamefont {Ren}}, \bibinfo {author} {\bibfnamefont {X.}~\bibnamefont {Fang}}, \bibinfo {author} {\bibfnamefont {J.}~\bibnamefont {Jang}}, \bibinfo {author} {\bibfnamefont {J.}~\bibnamefont {B{\"u}rger}}, \bibinfo {author} {\bibfnamefont {J.}~\bibnamefont {Rho}},\ and\ \bibinfo {author} {\bibfnamefont {S.~A.}\ \bibnamefont {Maier}},\ }\bibfield  {title} {\bibinfo {title} {Complex-amplitude metasurface-based orbital angular momentum holography in momentum space},\ }\href {https://doi.org/10.1038/s41565-020-0768-4} {\bibfield  {journal} {\bibinfo  {journal} {Nat. Nanotechnol.}\ }\textbf {\bibinfo {volume} {15}},\ \bibinfo {pages} {948} (\bibinfo {year} {2020})}\BibitemShut {NoStop}%
\bibitem [{\citenamefont {Bao}\ \emph {et~al.}(2022)\citenamefont {Bao}, \citenamefont {Nan}, \citenamefont {Yan}, \citenamefont {Yang}, \citenamefont {Qiu},\ and\ \citenamefont {Li}}]{Bao2022-mz}%
  \BibitemOpen
  \bibfield  {author} {\bibinfo {author} {\bibfnamefont {Y.}~\bibnamefont {Bao}}, \bibinfo {author} {\bibfnamefont {F.}~\bibnamefont {Nan}}, \bibinfo {author} {\bibfnamefont {J.}~\bibnamefont {Yan}}, \bibinfo {author} {\bibfnamefont {X.}~\bibnamefont {Yang}}, \bibinfo {author} {\bibfnamefont {C.-W.}\ \bibnamefont {Qiu}},\ and\ \bibinfo {author} {\bibfnamefont {B.}~\bibnamefont {Li}},\ }\bibfield  {title} {\bibinfo {title} {Observation of full-parameter {Jones} matrix in bilayer metasurface},\ }\href {https://doi.org/10.1038/s41467-022-35313-2} {\bibfield  {journal} {\bibinfo  {journal} {Nat. Commun.}\ }\textbf {\bibinfo {volume} {13}},\ \bibinfo {pages} {7550} (\bibinfo {year} {2022})}\BibitemShut {NoStop}%
\bibitem [{\citenamefont {Zaidi}\ \emph {et~al.}(2024)\citenamefont {Zaidi}, \citenamefont {Rubin}, \citenamefont {Meretska}, \citenamefont {Li}, \citenamefont {Dorrah}, \citenamefont {Park},\ and\ \citenamefont {Capasso}}]{Zaidi2024-lk}%
  \BibitemOpen
  \bibfield  {author} {\bibinfo {author} {\bibfnamefont {A.}~\bibnamefont {Zaidi}}, \bibinfo {author} {\bibfnamefont {N.~A.}\ \bibnamefont {Rubin}}, \bibinfo {author} {\bibfnamefont {M.~L.}\ \bibnamefont {Meretska}}, \bibinfo {author} {\bibfnamefont {L.~W.}\ \bibnamefont {Li}}, \bibinfo {author} {\bibfnamefont {A.~H.}\ \bibnamefont {Dorrah}}, \bibinfo {author} {\bibfnamefont {J.-S.}\ \bibnamefont {Park}},\ and\ \bibinfo {author} {\bibfnamefont {F.}~\bibnamefont {Capasso}},\ }\bibfield  {title} {\bibinfo {title} {Metasurface-enabled single-shot and complete mueller matrix imaging},\ }\href {https://doi.org/10.1038/s41566-024-01426-x} {\bibfield  {journal} {\bibinfo  {journal} {Nat. Photonics}\ ,\ \bibinfo {pages} {1}} (\bibinfo {year} {2024})}\BibitemShut {NoStop}%
\bibitem [{\citenamefont {Lee}\ \emph {et~al.}(2018)\citenamefont {Lee}, \citenamefont {Hong}, \citenamefont {Hwang}, \citenamefont {Moon}, \citenamefont {Kang}, \citenamefont {Jeon}, \citenamefont {Kim}, \citenamefont {Jeong},\ and\ \citenamefont {Lee}}]{Lee2018-eq}%
  \BibitemOpen
  \bibfield  {author} {\bibinfo {author} {\bibfnamefont {G.-Y.}\ \bibnamefont {Lee}}, \bibinfo {author} {\bibfnamefont {J.-Y.}\ \bibnamefont {Hong}}, \bibinfo {author} {\bibfnamefont {S.}~\bibnamefont {Hwang}}, \bibinfo {author} {\bibfnamefont {S.}~\bibnamefont {Moon}}, \bibinfo {author} {\bibfnamefont {H.}~\bibnamefont {Kang}}, \bibinfo {author} {\bibfnamefont {S.}~\bibnamefont {Jeon}}, \bibinfo {author} {\bibfnamefont {H.}~\bibnamefont {Kim}}, \bibinfo {author} {\bibfnamefont {J.-H.}\ \bibnamefont {Jeong}},\ and\ \bibinfo {author} {\bibfnamefont {B.}~\bibnamefont {Lee}},\ }\bibfield  {title} {\bibinfo {title} {Metasurface eyepiece for augmented reality},\ }\href {https://doi.org/10.1038/s41467-018-07011-5} {\bibfield  {journal} {\bibinfo  {journal} {Nat. Commun.}\ }\textbf {\bibinfo {volume} {9}},\ \bibinfo {pages} {4562} (\bibinfo {year} {2018})}\BibitemShut {NoStop}%
\bibitem [{\citenamefont {Joo}\ \emph {et~al.}(2020)\citenamefont {Joo}, \citenamefont {Kyoung}, \citenamefont {Esfandyarpour}, \citenamefont {Lee}, \citenamefont {Koo}, \citenamefont {Song}, \citenamefont {Kwon}, \citenamefont {Song}, \citenamefont {Bae}, \citenamefont {Jo}, \citenamefont {Kwon}, \citenamefont {Han}, \citenamefont {Kim}, \citenamefont {Hwang},\ and\ \citenamefont {Brongersma}}]{Joo2020-lp}%
  \BibitemOpen
  \bibfield  {author} {\bibinfo {author} {\bibfnamefont {W.-J.}\ \bibnamefont {Joo}}, \bibinfo {author} {\bibfnamefont {J.}~\bibnamefont {Kyoung}}, \bibinfo {author} {\bibfnamefont {M.}~\bibnamefont {Esfandyarpour}}, \bibinfo {author} {\bibfnamefont {S.-H.}\ \bibnamefont {Lee}}, \bibinfo {author} {\bibfnamefont {H.}~\bibnamefont {Koo}}, \bibinfo {author} {\bibfnamefont {S.}~\bibnamefont {Song}}, \bibinfo {author} {\bibfnamefont {Y.-N.}\ \bibnamefont {Kwon}}, \bibinfo {author} {\bibfnamefont {S.~H.}\ \bibnamefont {Song}}, \bibinfo {author} {\bibfnamefont {J.~C.}\ \bibnamefont {Bae}}, \bibinfo {author} {\bibfnamefont {A.}~\bibnamefont {Jo}}, \bibinfo {author} {\bibfnamefont {M.-J.}\ \bibnamefont {Kwon}}, \bibinfo {author} {\bibfnamefont {S.~H.}\ \bibnamefont {Han}}, \bibinfo {author} {\bibfnamefont {S.-H.}\ \bibnamefont {Kim}}, \bibinfo {author} {\bibfnamefont {S.}~\bibnamefont {Hwang}},\ and\ \bibinfo {author} {\bibfnamefont {M.~L.}\ \bibnamefont {Brongersma}},\ }\bibfield  {title} {\bibinfo {title}
  {Metasurface-driven {OLED} displays beyond 10,000 pixels per inch},\ }\href {https://doi.org/10.1126/science.abc8530} {\bibfield  {journal} {\bibinfo  {journal} {Science}\ }\textbf {\bibinfo {volume} {370}},\ \bibinfo {pages} {459} (\bibinfo {year} {2020})}\BibitemShut {NoStop}%
\bibitem [{\citenamefont {Li}\ \emph {et~al.}(2021)\citenamefont {Li}, \citenamefont {Lin}, \citenamefont {Huang}, \citenamefont {Park}, \citenamefont {Chen}, \citenamefont {Shi}, \citenamefont {Qiu}, \citenamefont {Cheng},\ and\ \citenamefont {Capasso}}]{Li2021-gi}%
  \BibitemOpen
  \bibfield  {author} {\bibinfo {author} {\bibfnamefont {Z.}~\bibnamefont {Li}}, \bibinfo {author} {\bibfnamefont {P.}~\bibnamefont {Lin}}, \bibinfo {author} {\bibfnamefont {Y.-W.}\ \bibnamefont {Huang}}, \bibinfo {author} {\bibfnamefont {J.-S.}\ \bibnamefont {Park}}, \bibinfo {author} {\bibfnamefont {W.~T.}\ \bibnamefont {Chen}}, \bibinfo {author} {\bibfnamefont {Z.}~\bibnamefont {Shi}}, \bibinfo {author} {\bibfnamefont {C.-W.}\ \bibnamefont {Qiu}}, \bibinfo {author} {\bibfnamefont {J.-X.}\ \bibnamefont {Cheng}},\ and\ \bibinfo {author} {\bibfnamefont {F.}~\bibnamefont {Capasso}},\ }\bibfield  {title} {\bibinfo {title} {Meta-optics achieves {RGB}-achromatic focusing for virtual reality},\ }\href {https://doi.org/10.1126/sciadv.abe4458} {\bibfield  {journal} {\bibinfo  {journal} {Sci. Adv.}\ }\textbf {\bibinfo {volume} {7}},\ \bibinfo {pages} {eabe4458} (\bibinfo {year} {2021})}\BibitemShut {NoStop}%
\bibitem [{\citenamefont {Li}\ \emph {et~al.}(2022)\citenamefont {Li}, \citenamefont {Pestourie}, \citenamefont {Park}, \citenamefont {Huang}, \citenamefont {Johnson},\ and\ \citenamefont {Capasso}}]{Li2022-zu}%
  \BibitemOpen
  \bibfield  {author} {\bibinfo {author} {\bibfnamefont {Z.}~\bibnamefont {Li}}, \bibinfo {author} {\bibfnamefont {R.}~\bibnamefont {Pestourie}}, \bibinfo {author} {\bibfnamefont {J.-S.}\ \bibnamefont {Park}}, \bibinfo {author} {\bibfnamefont {Y.-W.}\ \bibnamefont {Huang}}, \bibinfo {author} {\bibfnamefont {S.~G.}\ \bibnamefont {Johnson}},\ and\ \bibinfo {author} {\bibfnamefont {F.}~\bibnamefont {Capasso}},\ }\bibfield  {title} {\bibinfo {title} {Inverse design enables large-scale high-performance meta-optics reshaping virtual reality},\ }\href {https://doi.org/10.1038/s41467-022-29973-3} {\bibfield  {journal} {\bibinfo  {journal} {Nat. Commun.}\ }\textbf {\bibinfo {volume} {13}},\ \bibinfo {pages} {2409} (\bibinfo {year} {2022})}\BibitemShut {NoStop}%
\bibitem [{\citenamefont {Neshev}\ and\ \citenamefont {Miroshnichenko}(2023)}]{neshev2023enabling}%
  \BibitemOpen
  \bibfield  {author} {\bibinfo {author} {\bibfnamefont {D.~N.}\ \bibnamefont {Neshev}}\ and\ \bibinfo {author} {\bibfnamefont {A.~E.}\ \bibnamefont {Miroshnichenko}},\ }\bibfield  {title} {\bibinfo {title} {Enabling smart vision with metasurfaces},\ }\href {https://www.nature.com/articles/s41566-022-01126-4} {\bibfield  {journal} {\bibinfo  {journal} {Nat. Photonics}\ }\textbf {\bibinfo {volume} {17}},\ \bibinfo {pages} {26} (\bibinfo {year} {2023})}\BibitemShut {NoStop}%
\bibitem [{\citenamefont {Gopakumar}\ \emph {et~al.}(2024)\citenamefont {Gopakumar}, \citenamefont {Lee}, \citenamefont {Choi}, \citenamefont {Chao}, \citenamefont {Peng}, \citenamefont {Kim},\ and\ \citenamefont {Wetzstein}}]{Gopakumar2024-ad}%
  \BibitemOpen
  \bibfield  {author} {\bibinfo {author} {\bibfnamefont {M.}~\bibnamefont {Gopakumar}}, \bibinfo {author} {\bibfnamefont {G.-Y.}\ \bibnamefont {Lee}}, \bibinfo {author} {\bibfnamefont {S.}~\bibnamefont {Choi}}, \bibinfo {author} {\bibfnamefont {B.}~\bibnamefont {Chao}}, \bibinfo {author} {\bibfnamefont {Y.}~\bibnamefont {Peng}}, \bibinfo {author} {\bibfnamefont {J.}~\bibnamefont {Kim}},\ and\ \bibinfo {author} {\bibfnamefont {G.}~\bibnamefont {Wetzstein}},\ }\bibfield  {title} {\bibinfo {title} {Full-colour {3D} holographic augmented-reality displays with metasurface waveguides},\ }\href {https://doi.org/10.1038/s41586-024-07386-0} {\bibfield  {journal} {\bibinfo  {journal} {Nature}\ }\textbf {\bibinfo {volume} {629}},\ \bibinfo {pages} {791} (\bibinfo {year} {2024})}\BibitemShut {NoStop}%
\bibitem [{\citenamefont {Richardson}\ \emph {et~al.}(2013)\citenamefont {Richardson}, \citenamefont {Fini},\ and\ \citenamefont {Nelson}}]{Richardson2013-qw}%
  \BibitemOpen
  \bibfield  {author} {\bibinfo {author} {\bibfnamefont {D.~J.}\ \bibnamefont {Richardson}}, \bibinfo {author} {\bibfnamefont {J.~M.}\ \bibnamefont {Fini}},\ and\ \bibinfo {author} {\bibfnamefont {L.~E.}\ \bibnamefont {Nelson}},\ }\bibfield  {title} {\bibinfo {title} {Space-division multiplexing in optical fibres},\ }\href {https://doi.org/10.1038/nphoton.2013.94} {\bibfield  {journal} {\bibinfo  {journal} {Nat. Photonics}\ }\textbf {\bibinfo {volume} {7}},\ \bibinfo {pages} {354} (\bibinfo {year} {2013})}\BibitemShut {NoStop}%
\bibitem [{\citenamefont {Mizuno}\ \emph {et~al.}(2016)\citenamefont {Mizuno}, \citenamefont {Takara}, \citenamefont {Sano},\ and\ \citenamefont {Miyamoto}}]{Mizuno2016-na}%
  \BibitemOpen
  \bibfield  {author} {\bibinfo {author} {\bibfnamefont {T.}~\bibnamefont {Mizuno}}, \bibinfo {author} {\bibfnamefont {H.}~\bibnamefont {Takara}}, \bibinfo {author} {\bibfnamefont {A.}~\bibnamefont {Sano}},\ and\ \bibinfo {author} {\bibfnamefont {Y.}~\bibnamefont {Miyamoto}},\ }\bibfield  {title} {\bibinfo {title} {Dense space-division multiplexed transmission systems using multi-core and multi-mode fiber},\ }\href {https://www.osapublishing.org/abstract.cfm?uri=jlt-34-2-582} {\bibfield  {journal} {\bibinfo  {journal} {J. Lightwave Technol.}\ }\textbf {\bibinfo {volume} {34}},\ \bibinfo {pages} {582} (\bibinfo {year} {2016})}\BibitemShut {NoStop}%
\bibitem [{\citenamefont {Puttnam}\ \emph {et~al.}(2021)\citenamefont {Puttnam}, \citenamefont {Rademacher},\ and\ \citenamefont {Lu{\'\i}s}}]{puttnam2021space}%
  \BibitemOpen
  \bibfield  {author} {\bibinfo {author} {\bibfnamefont {B.~J.}\ \bibnamefont {Puttnam}}, \bibinfo {author} {\bibfnamefont {G.}~\bibnamefont {Rademacher}},\ and\ \bibinfo {author} {\bibfnamefont {R.~S.}\ \bibnamefont {Lu{\'\i}s}},\ }\bibfield  {title} {\bibinfo {title} {Space-division multiplexing for optical fiber communications},\ }\href {https://doi.org/10.1364/OPTICA.427631} {\bibfield  {journal} {\bibinfo  {journal} {Optica}\ }\textbf {\bibinfo {volume} {8}},\ \bibinfo {pages} {1186} (\bibinfo {year} {2021})}\BibitemShut {NoStop}%
\bibitem [{\citenamefont {Kikuchi}(2016)}]{Kikuchi2016-pd}%
  \BibitemOpen
  \bibfield  {author} {\bibinfo {author} {\bibfnamefont {K.}~\bibnamefont {Kikuchi}},\ }\bibfield  {title} {\bibinfo {title} {Fundamentals of coherent optical fiber communications},\ }\href {https://doi.org/10.1109/JLT.2015.2463719} {\bibfield  {journal} {\bibinfo  {journal} {J. Lightwave Technol.}\ }\textbf {\bibinfo {volume} {34}},\ \bibinfo {pages} {157} (\bibinfo {year} {2016})}\BibitemShut {NoStop}%
\bibitem [{\citenamefont {Dong}\ \emph {et~al.}(2014)\citenamefont {Dong}, \citenamefont {Liu}, \citenamefont {Chandrasekhar}, \citenamefont {Buhl}, \citenamefont {Aroca},\ and\ \citenamefont {Chen}}]{Dong2014-qf}%
  \BibitemOpen
  \bibfield  {author} {\bibinfo {author} {\bibfnamefont {P.}~\bibnamefont {Dong}}, \bibinfo {author} {\bibfnamefont {X.}~\bibnamefont {Liu}}, \bibinfo {author} {\bibfnamefont {S.}~\bibnamefont {Chandrasekhar}}, \bibinfo {author} {\bibfnamefont {L.~L.}\ \bibnamefont {Buhl}}, \bibinfo {author} {\bibfnamefont {R.}~\bibnamefont {Aroca}},\ and\ \bibinfo {author} {\bibfnamefont {Y.~K.}\ \bibnamefont {Chen}},\ }\bibfield  {title} {\bibinfo {title} {Monolithic silicon photonic integrated circuits for compact 100 +{Gb/s} coherent optical receivers and transmitters},\ }\href {https://doi.org/10.1109/JSTQE.2013.2295181} {\bibfield  {journal} {\bibinfo  {journal} {IEEE J. Sel. Top. Quantum Electron.}\ }\textbf {\bibinfo {volume} {20}},\ \bibinfo {pages} {6100108} (\bibinfo {year} {2014})}\BibitemShut {NoStop}%
\bibitem [{\citenamefont {Ogiso}\ \emph {et~al.}(2017)\citenamefont {Ogiso}, \citenamefont {Ozaki}, \citenamefont {Ueda}, \citenamefont {Kashio}, \citenamefont {Kikuchi}, \citenamefont {Yamada}, \citenamefont {Tanobe}, \citenamefont {Kanazawa}, \citenamefont {Yamazaki}, \citenamefont {Ohiso}, \citenamefont {Fujii},\ and\ \citenamefont {Kohtoku}}]{Ogiso2017-hb}%
  \BibitemOpen
  \bibfield  {author} {\bibinfo {author} {\bibfnamefont {Y.}~\bibnamefont {Ogiso}}, \bibinfo {author} {\bibfnamefont {J.}~\bibnamefont {Ozaki}}, \bibinfo {author} {\bibfnamefont {Y.}~\bibnamefont {Ueda}}, \bibinfo {author} {\bibfnamefont {N.}~\bibnamefont {Kashio}}, \bibinfo {author} {\bibfnamefont {N.}~\bibnamefont {Kikuchi}}, \bibinfo {author} {\bibfnamefont {E.}~\bibnamefont {Yamada}}, \bibinfo {author} {\bibfnamefont {H.}~\bibnamefont {Tanobe}}, \bibinfo {author} {\bibfnamefont {S.}~\bibnamefont {Kanazawa}}, \bibinfo {author} {\bibfnamefont {H.}~\bibnamefont {Yamazaki}}, \bibinfo {author} {\bibfnamefont {Y.}~\bibnamefont {Ohiso}}, \bibinfo {author} {\bibfnamefont {T.}~\bibnamefont {Fujii}},\ and\ \bibinfo {author} {\bibfnamefont {M.}~\bibnamefont {Kohtoku}},\ }\bibfield  {title} {\bibinfo {title} {Over 67 {GHz} bandwidth and 1.5 {V} {V}$\pi$ {InP}-based optical {IQ} modulator with n-i-p-n heterostructure},\ }\href {https://opg.optica.org/abstract.cfm?uri=jlt-35-8-1450} {\bibfield  {journal} {\bibinfo
  {journal} {J. Lightwave Technol.}\ }\textbf {\bibinfo {volume} {35}},\ \bibinfo {pages} {1450} (\bibinfo {year} {2017})}\BibitemShut {NoStop}%
\bibitem [{\citenamefont {Yagi}\ \emph {et~al.}(2018)\citenamefont {Yagi}, \citenamefont {Kaneko}, \citenamefont {Kono}, \citenamefont {Yoneda}, \citenamefont {Uesaka}, \citenamefont {Ekawa}, \citenamefont {Takechi},\ and\ \citenamefont {Shoji}}]{Yagi2018-do}%
  \BibitemOpen
  \bibfield  {author} {\bibinfo {author} {\bibfnamefont {H.}~\bibnamefont {Yagi}}, \bibinfo {author} {\bibfnamefont {T.}~\bibnamefont {Kaneko}}, \bibinfo {author} {\bibfnamefont {N.}~\bibnamefont {Kono}}, \bibinfo {author} {\bibfnamefont {Y.}~\bibnamefont {Yoneda}}, \bibinfo {author} {\bibfnamefont {K.}~\bibnamefont {Uesaka}}, \bibinfo {author} {\bibfnamefont {M.}~\bibnamefont {Ekawa}}, \bibinfo {author} {\bibfnamefont {M.}~\bibnamefont {Takechi}},\ and\ \bibinfo {author} {\bibfnamefont {H.}~\bibnamefont {Shoji}},\ }\bibfield  {title} {\bibinfo {title} {{InP}-based monolithically integrated photonic devices for digital coherent transmission},\ }\href {https://doi.org/10.1109/JSTQE.2017.2725445} {\bibfield  {journal} {\bibinfo  {journal} {IEEE J. Sel. Top. Quantum Electron.}\ }\textbf {\bibinfo {volume} {24}},\ \bibinfo {pages} {6100411} (\bibinfo {year} {2018})}\BibitemShut {NoStop}%
\bibitem [{\citenamefont {Porto}\ \emph {et~al.}(2022)\citenamefont {Porto}, \citenamefont {Chitgarha}, \citenamefont {Leung}, \citenamefont {Maher}, \citenamefont {Going}, \citenamefont {Wolf}, \citenamefont {Studenkov}, \citenamefont {Zhang}, \citenamefont {Hodaei}, \citenamefont {Frost}, \citenamefont {Tsai}, \citenamefont {Buggaveeti}, \citenamefont {Rashidinejad}, \citenamefont {Yekani}, \citenamefont {Nejad}, \citenamefont {Olmedo}, \citenamefont {Kerns}, \citenamefont {Diniz}, \citenamefont {Pavinski}, \citenamefont {Brigham}, \citenamefont {Foo}, \citenamefont {Al-Khateeb}, \citenamefont {Koenig}, \citenamefont {Samra}, \citenamefont {Missey}, \citenamefont {Dominic}, \citenamefont {Sun}, \citenamefont {Sanders}, \citenamefont {Osenbach}, \citenamefont {Corzine}, \citenamefont {Evans}, \citenamefont {Lal},\ and\ \citenamefont {Ziari}}]{Porto2022-nz}%
  \BibitemOpen
  \bibfield  {author} {\bibinfo {author} {\bibfnamefont {S.}~\bibnamefont {Porto}}, \bibinfo {author} {\bibfnamefont {M.}~\bibnamefont {Chitgarha}}, \bibinfo {author} {\bibfnamefont {I.}~\bibnamefont {Leung}}, \bibinfo {author} {\bibfnamefont {R.}~\bibnamefont {Maher}}, \bibinfo {author} {\bibfnamefont {R.}~\bibnamefont {Going}}, \bibinfo {author} {\bibfnamefont {S.}~\bibnamefont {Wolf}}, \bibinfo {author} {\bibfnamefont {P.}~\bibnamefont {Studenkov}}, \bibinfo {author} {\bibfnamefont {J.}~\bibnamefont {Zhang}}, \bibinfo {author} {\bibfnamefont {H.}~\bibnamefont {Hodaei}}, \bibinfo {author} {\bibfnamefont {T.}~\bibnamefont {Frost}}, \bibinfo {author} {\bibfnamefont {H.-S.}\ \bibnamefont {Tsai}}, \bibinfo {author} {\bibfnamefont {S.}~\bibnamefont {Buggaveeti}}, \bibinfo {author} {\bibfnamefont {A.}~\bibnamefont {Rashidinejad}}, \bibinfo {author} {\bibfnamefont {A.}~\bibnamefont {Yekani}}, \bibinfo {author} {\bibfnamefont {R.~M.}\ \bibnamefont {Nejad}}, \bibinfo {author} {\bibfnamefont {M.~I.}\ \bibnamefont
  {Olmedo}}, \bibinfo {author} {\bibfnamefont {S.}~\bibnamefont {Kerns}}, \bibinfo {author} {\bibfnamefont {J.}~\bibnamefont {Diniz}}, \bibinfo {author} {\bibfnamefont {D.}~\bibnamefont {Pavinski}}, \bibinfo {author} {\bibfnamefont {R.}~\bibnamefont {Brigham}}, \bibinfo {author} {\bibfnamefont {B.}~\bibnamefont {Foo}}, \bibinfo {author} {\bibfnamefont {M.}~\bibnamefont {Al-Khateeb}}, \bibinfo {author} {\bibfnamefont {S.}~\bibnamefont {Koenig}}, \bibinfo {author} {\bibfnamefont {P.}~\bibnamefont {Samra}}, \bibinfo {author} {\bibfnamefont {M.}~\bibnamefont {Missey}}, \bibinfo {author} {\bibfnamefont {V.}~\bibnamefont {Dominic}}, \bibinfo {author} {\bibfnamefont {H.}~\bibnamefont {Sun}}, \bibinfo {author} {\bibfnamefont {S.}~\bibnamefont {Sanders}}, \bibinfo {author} {\bibfnamefont {J.}~\bibnamefont {Osenbach}}, \bibinfo {author} {\bibfnamefont {S.}~\bibnamefont {Corzine}}, \bibinfo {author} {\bibfnamefont {P.}~\bibnamefont {Evans}}, \bibinfo {author} {\bibfnamefont {V.}~\bibnamefont {Lal}},\ and\ \bibinfo
  {author} {\bibfnamefont {M.}~\bibnamefont {Ziari}},\ }\bibfield  {title} {\bibinfo {title} {Demonstration of a 2 $\times$ 800 {Gb/s/wave} coherent optical engine based on an {InP} monolithic {PIC}},\ }\href {https://opg.optica.org/abstract.cfm?uri=jlt-40-3-664} {\bibfield  {journal} {\bibinfo  {journal} {J. Lightwave Technol.}\ }\textbf {\bibinfo {volume} {40}},\ \bibinfo {pages} {664} (\bibinfo {year} {2022})}\BibitemShut {NoStop}%
\bibitem [{\citenamefont {Xu}\ \emph {et~al.}(2020)\citenamefont {Xu}, \citenamefont {He}, \citenamefont {Zhang}, \citenamefont {Jian}, \citenamefont {Pan}, \citenamefont {Liu}, \citenamefont {Chen}, \citenamefont {Meng}, \citenamefont {Chen}, \citenamefont {Li}, \citenamefont {Xiao}, \citenamefont {Yu}, \citenamefont {Yu},\ and\ \citenamefont {Cai}}]{Xu2020-pm}%
  \BibitemOpen
  \bibfield  {author} {\bibinfo {author} {\bibfnamefont {M.}~\bibnamefont {Xu}}, \bibinfo {author} {\bibfnamefont {M.}~\bibnamefont {He}}, \bibinfo {author} {\bibfnamefont {H.}~\bibnamefont {Zhang}}, \bibinfo {author} {\bibfnamefont {J.}~\bibnamefont {Jian}}, \bibinfo {author} {\bibfnamefont {Y.}~\bibnamefont {Pan}}, \bibinfo {author} {\bibfnamefont {X.}~\bibnamefont {Liu}}, \bibinfo {author} {\bibfnamefont {L.}~\bibnamefont {Chen}}, \bibinfo {author} {\bibfnamefont {X.}~\bibnamefont {Meng}}, \bibinfo {author} {\bibfnamefont {H.}~\bibnamefont {Chen}}, \bibinfo {author} {\bibfnamefont {Z.}~\bibnamefont {Li}}, \bibinfo {author} {\bibfnamefont {X.}~\bibnamefont {Xiao}}, \bibinfo {author} {\bibfnamefont {S.}~\bibnamefont {Yu}}, \bibinfo {author} {\bibfnamefont {S.}~\bibnamefont {Yu}},\ and\ \bibinfo {author} {\bibfnamefont {X.}~\bibnamefont {Cai}},\ }\bibfield  {title} {\bibinfo {title} {High-performance coherent optical modulators based on thin-film lithium niobate platform},\ }\href
  {https://doi.org/10.1038/s41467-020-17806-0} {\bibfield  {journal} {\bibinfo  {journal} {Nat. Commun.}\ }\textbf {\bibinfo {volume} {11}},\ \bibinfo {pages} {3911} (\bibinfo {year} {2020})}\BibitemShut {NoStop}%
\bibitem [{\citenamefont {Marchetti}\ \emph {et~al.}(2019)\citenamefont {Marchetti}, \citenamefont {Lacava}, \citenamefont {Carroll}, \citenamefont {Gradkowski},\ and\ \citenamefont {Minzioni}}]{Marchetti2019-pv}%
  \BibitemOpen
  \bibfield  {author} {\bibinfo {author} {\bibfnamefont {R.}~\bibnamefont {Marchetti}}, \bibinfo {author} {\bibfnamefont {C.}~\bibnamefont {Lacava}}, \bibinfo {author} {\bibfnamefont {L.}~\bibnamefont {Carroll}}, \bibinfo {author} {\bibfnamefont {K.}~\bibnamefont {Gradkowski}},\ and\ \bibinfo {author} {\bibfnamefont {P.}~\bibnamefont {Minzioni}},\ }\bibfield  {title} {\bibinfo {title} {Coupling strategies for silicon photonics integrated chips [invited]},\ }\href {https://doi.org/10.1364/PRJ.7.000201} {\bibfield  {journal} {\bibinfo  {journal} {Photon. Res.}\ }\textbf {\bibinfo {volume} {7}},\ \bibinfo {pages} {201} (\bibinfo {year} {2019})}\BibitemShut {NoStop}%
\bibitem [{\citenamefont {Oh}\ \emph {et~al.}(2022)\citenamefont {Oh}, \citenamefont {Li}, \citenamefont {Yang}, \citenamefont {Chen}, \citenamefont {Li}, \citenamefont {Dainese},\ and\ \citenamefont {Capasso}}]{Oh2022-tp}%
  \BibitemOpen
  \bibfield  {author} {\bibinfo {author} {\bibfnamefont {J.}~\bibnamefont {Oh}}, \bibinfo {author} {\bibfnamefont {K.}~\bibnamefont {Li}}, \bibinfo {author} {\bibfnamefont {J.}~\bibnamefont {Yang}}, \bibinfo {author} {\bibfnamefont {W.~T.}\ \bibnamefont {Chen}}, \bibinfo {author} {\bibfnamefont {M.-J.}\ \bibnamefont {Li}}, \bibinfo {author} {\bibfnamefont {P.}~\bibnamefont {Dainese}},\ and\ \bibinfo {author} {\bibfnamefont {F.}~\bibnamefont {Capasso}},\ }\bibfield  {title} {\bibinfo {title} {Adjoint-optimized metasurfaces for compact mode-division multiplexing},\ }\href {https://doi.org/10.1021/acsphotonics.1c01744} {\bibfield  {journal} {\bibinfo  {journal} {ACS Photonics}\ }\textbf {\bibinfo {volume} {9}},\ \bibinfo {pages} {929} (\bibinfo {year} {2022})}\BibitemShut {NoStop}%
\bibitem [{\citenamefont {Oh}\ \emph {et~al.}(2024)\citenamefont {Oh}, \citenamefont {Yang}, \citenamefont {Marra}, \citenamefont {Dorrah}, \citenamefont {Palmieri}, \citenamefont {Dainese},\ and\ \citenamefont {Capasso}}]{Oh2024-ux}%
  \BibitemOpen
  \bibfield  {author} {\bibinfo {author} {\bibfnamefont {J.}~\bibnamefont {Oh}}, \bibinfo {author} {\bibfnamefont {J.}~\bibnamefont {Yang}}, \bibinfo {author} {\bibfnamefont {L.}~\bibnamefont {Marra}}, \bibinfo {author} {\bibfnamefont {A.~H.}\ \bibnamefont {Dorrah}}, \bibinfo {author} {\bibfnamefont {A.}~\bibnamefont {Palmieri}}, \bibinfo {author} {\bibfnamefont {P.}~\bibnamefont {Dainese}},\ and\ \bibinfo {author} {\bibfnamefont {F.}~\bibnamefont {Capasso}},\ }\bibfield  {title} {\bibinfo {title} {Metasurfaces for free-space coupling to multicore fibers},\ }\href {https://doi.org/10.1109/JLT.2023.3335334} {\bibfield  {journal} {\bibinfo  {journal} {J. Lightwave Technol.}\ }\textbf {\bibinfo {volume} {42}},\ \bibinfo {pages} {2385} (\bibinfo {year} {2024})}\BibitemShut {NoStop}%
\bibitem [{\citenamefont {Soma}\ \emph {et~al.}(2023)\citenamefont {Soma}, \citenamefont {Nomoto}, \citenamefont {Umezawa}, \citenamefont {Yoshida}, \citenamefont {Nakano},\ and\ \citenamefont {Tanemura}}]{Soma2023-fx}%
  \BibitemOpen
  \bibfield  {author} {\bibinfo {author} {\bibfnamefont {G.}~\bibnamefont {Soma}}, \bibinfo {author} {\bibfnamefont {Y.}~\bibnamefont {Nomoto}}, \bibinfo {author} {\bibfnamefont {T.}~\bibnamefont {Umezawa}}, \bibinfo {author} {\bibfnamefont {Y.}~\bibnamefont {Yoshida}}, \bibinfo {author} {\bibfnamefont {Y.}~\bibnamefont {Nakano}},\ and\ \bibinfo {author} {\bibfnamefont {T.}~\bibnamefont {Tanemura}},\ }\bibfield  {title} {\bibinfo {title} {Compact and scalable polarimetric self-coherent receiver using a dielectric metasurface},\ }\href {https://doi.org/10.1364/optica.484318} {\bibfield  {journal} {\bibinfo  {journal} {Optica}\ }\textbf {\bibinfo {volume} {10}},\ \bibinfo {pages} {604} (\bibinfo {year} {2023})}\BibitemShut {NoStop}%
\bibitem [{\citenamefont {Komatsu}\ \emph {et~al.}(2024)\citenamefont {Komatsu}, \citenamefont {Soma}, \citenamefont {Ishimura}, \citenamefont {Takahashi}, \citenamefont {Tsuritani}, \citenamefont {Suzuki}, \citenamefont {Nakano},\ and\ \citenamefont {Tanemura}}]{Komatsu2024-pi}%
  \BibitemOpen
  \bibfield  {author} {\bibinfo {author} {\bibfnamefont {K.}~\bibnamefont {Komatsu}}, \bibinfo {author} {\bibfnamefont {G.}~\bibnamefont {Soma}}, \bibinfo {author} {\bibfnamefont {S.}~\bibnamefont {Ishimura}}, \bibinfo {author} {\bibfnamefont {H.}~\bibnamefont {Takahashi}}, \bibinfo {author} {\bibfnamefont {T.}~\bibnamefont {Tsuritani}}, \bibinfo {author} {\bibfnamefont {M.}~\bibnamefont {Suzuki}}, \bibinfo {author} {\bibfnamefont {Y.}~\bibnamefont {Nakano}},\ and\ \bibinfo {author} {\bibfnamefont {T.}~\bibnamefont {Tanemura}},\ }\bibfield  {title} {\bibinfo {title} {Scalable multi-core dual-polarization coherent receiver using a metasurface optical hybrid},\ }\href {https://doi.org/10.1109/jlt.2024.3374336} {\bibfield  {journal} {\bibinfo  {journal} {J. Lightwave Technol.}\ }\textbf {\bibinfo {volume} {42}},\ \bibinfo {pages} {4013} (\bibinfo {year} {2024})}\BibitemShut {NoStop}%
\bibitem [{\citenamefont {Komatsu}\ \emph {et~al.}()\citenamefont {Komatsu}, \citenamefont {Ishimura}, \citenamefont {Ren}, \citenamefont {Soma}, \citenamefont {Takahashi}, \citenamefont {Suzuki}, \citenamefont {Nakano},\ and\ \citenamefont {Tanemura}}]{Komatsu2024-OFC}%
  \BibitemOpen
  \bibfield  {author} {\bibinfo {author} {\bibfnamefont {K.}~\bibnamefont {Komatsu}}, \bibinfo {author} {\bibfnamefont {S.}~\bibnamefont {Ishimura}}, \bibinfo {author} {\bibfnamefont {C.}~\bibnamefont {Ren}}, \bibinfo {author} {\bibfnamefont {G.}~\bibnamefont {Soma}}, \bibinfo {author} {\bibfnamefont {H.}~\bibnamefont {Takahashi}}, \bibinfo {author} {\bibfnamefont {M.}~\bibnamefont {Suzuki}}, \bibinfo {author} {\bibfnamefont {Y.}~\bibnamefont {Nakano}},\ and\ \bibinfo {author} {\bibfnamefont {T.}~\bibnamefont {Tanemura}},\ }\bibfield  {title} {\bibinfo {title} {Metasurface-based coherent receiver insensitive to {LO} polarization},\ }in\ \href {https://doi.org/10.1364/ofc.2024.th4b.2} {\emph {\bibinfo {booktitle} {Optical Fiber Communications Conference ({OFC}) 2024}}},\ p.\ \bibinfo {pages} {Th4B.2}\BibitemShut {NoStop}%
\bibitem [{\citenamefont {Ren}\ \emph {et~al.}(2024)\citenamefont {Ren}, \citenamefont {Komatsu}, \citenamefont {Soma}, \citenamefont {Nakano},\ and\ \citenamefont {Tanemura}}]{Ren2024-wz}%
  \BibitemOpen
  \bibfield  {author} {\bibinfo {author} {\bibfnamefont {C.}~\bibnamefont {Ren}}, \bibinfo {author} {\bibfnamefont {K.}~\bibnamefont {Komatsu}}, \bibinfo {author} {\bibfnamefont {G.}~\bibnamefont {Soma}}, \bibinfo {author} {\bibfnamefont {Y.}~\bibnamefont {Nakano}},\ and\ \bibinfo {author} {\bibfnamefont {T.}~\bibnamefont {Tanemura}},\ }\bibfield  {title} {\bibinfo {title} {Metasurface-based functional optical splitter for a spatially parallelized dual-polarization coherent modulator},\ }\href {https://doi.org/10.1364/OL.541473} {\bibfield  {journal} {\bibinfo  {journal} {Opt. Lett.}\ }\textbf {\bibinfo {volume} {49}},\ \bibinfo {pages} {7238} (\bibinfo {year} {2024})}\BibitemShut {NoStop}%
\bibitem [{\citenamefont {Nozaki}\ \emph {et~al.}(2016)\citenamefont {Nozaki}, \citenamefont {Matsuo}, \citenamefont {Fujii}, \citenamefont {Takeda}, \citenamefont {Ono}, \citenamefont {Shakoor}, \citenamefont {Kuramochi},\ and\ \citenamefont {Notomi}}]{Nozaki2016-yj}%
  \BibitemOpen
  \bibfield  {author} {\bibinfo {author} {\bibfnamefont {K.}~\bibnamefont {Nozaki}}, \bibinfo {author} {\bibfnamefont {S.}~\bibnamefont {Matsuo}}, \bibinfo {author} {\bibfnamefont {T.}~\bibnamefont {Fujii}}, \bibinfo {author} {\bibfnamefont {K.}~\bibnamefont {Takeda}}, \bibinfo {author} {\bibfnamefont {M.}~\bibnamefont {Ono}}, \bibinfo {author} {\bibfnamefont {A.}~\bibnamefont {Shakoor}}, \bibinfo {author} {\bibfnamefont {E.}~\bibnamefont {Kuramochi}},\ and\ \bibinfo {author} {\bibfnamefont {M.}~\bibnamefont {Notomi}},\ }\bibfield  {title} {\bibinfo {title} {Photonic-crystal nano-photodetector with ultrasmall capacitance for on-chip light-to-voltage conversion without an amplifier},\ }\href {https://doi.org/10.1364/optica.3.000483} {\bibfield  {journal} {\bibinfo  {journal} {Optica}\ }\textbf {\bibinfo {volume} {3}},\ \bibinfo {pages} {483} (\bibinfo {year} {2016})}\BibitemShut {NoStop}%
\bibitem [{\citenamefont {Shen}\ \emph {et~al.}(2016)\citenamefont {Shen}, \citenamefont {Jiao}, \citenamefont {Yao}, \citenamefont {Cao}, \citenamefont {van Engelen}, \citenamefont {Roelkens}, \citenamefont {Smit},\ and\ \citenamefont {van~der Tol}}]{Shen2016-ve}%
  \BibitemOpen
  \bibfield  {author} {\bibinfo {author} {\bibfnamefont {L.}~\bibnamefont {Shen}}, \bibinfo {author} {\bibfnamefont {Y.}~\bibnamefont {Jiao}}, \bibinfo {author} {\bibfnamefont {W.}~\bibnamefont {Yao}}, \bibinfo {author} {\bibfnamefont {Z.}~\bibnamefont {Cao}}, \bibinfo {author} {\bibfnamefont {J.~P.}\ \bibnamefont {van Engelen}}, \bibinfo {author} {\bibfnamefont {G.}~\bibnamefont {Roelkens}}, \bibinfo {author} {\bibfnamefont {M.~K.}\ \bibnamefont {Smit}},\ and\ \bibinfo {author} {\bibfnamefont {J.~J. G.~M.}\ \bibnamefont {van~der Tol}},\ }\bibfield  {title} {\bibinfo {title} {High-bandwidth uni-traveling carrier waveguide photodetector on an {InP}-membrane-on-silicon platform},\ }\href {https://doi.org/10.1364/OE.24.008290} {\bibfield  {journal} {\bibinfo  {journal} {Opt. Express}\ }\textbf {\bibinfo {volume} {24}},\ \bibinfo {pages} {8290} (\bibinfo {year} {2016})}\BibitemShut {NoStop}%
\bibitem [{\citenamefont {Tyo}\ \emph {et~al.}(2006)\citenamefont {Tyo}, \citenamefont {Goldstein}, \citenamefont {Chenault},\ and\ \citenamefont {Shaw}}]{Tyo2006-bi}%
  \BibitemOpen
  \bibfield  {author} {\bibinfo {author} {\bibfnamefont {J.~S.}\ \bibnamefont {Tyo}}, \bibinfo {author} {\bibfnamefont {D.~L.}\ \bibnamefont {Goldstein}}, \bibinfo {author} {\bibfnamefont {D.~B.}\ \bibnamefont {Chenault}},\ and\ \bibinfo {author} {\bibfnamefont {J.~A.}\ \bibnamefont {Shaw}},\ }\bibfield  {title} {\bibinfo {title} {Review of passive imaging polarimetry for remote sensing applications},\ }\href {https://doi.org/10.1364/ao.45.005453} {\bibfield  {journal} {\bibinfo  {journal} {Appl. Opt.}\ }\textbf {\bibinfo {volume} {45}},\ \bibinfo {pages} {5453} (\bibinfo {year} {2006})}\BibitemShut {NoStop}%
\bibitem [{\citenamefont {Tanemura}\ \emph {et~al.}(2020)\citenamefont {Tanemura}, \citenamefont {Suganuma},\ and\ \citenamefont {Nakano}}]{Tanemura2020-nm}%
  \BibitemOpen
  \bibfield  {author} {\bibinfo {author} {\bibfnamefont {T.}~\bibnamefont {Tanemura}}, \bibinfo {author} {\bibfnamefont {T.}~\bibnamefont {Suganuma}},\ and\ \bibinfo {author} {\bibfnamefont {Y.}~\bibnamefont {Nakano}},\ }\bibfield  {title} {\bibinfo {title} {Sensitivity analysis of photonic integrated direct-detection {Stokes}-vector receiver},\ }\href {https://doi.org/10.1109/JLT.2019.2952980} {\bibfield  {journal} {\bibinfo  {journal} {J. Lightwave Technol.}\ }\textbf {\bibinfo {volume} {38}},\ \bibinfo {pages} {447} (\bibinfo {year} {2020})}\BibitemShut {NoStop}%
\bibitem [{\citenamefont {Soma}\ \emph {et~al.}(2022)\citenamefont {Soma}, \citenamefont {Yanwachirakul}, \citenamefont {Miyazaki}, \citenamefont {Kato}, \citenamefont {Onodera}, \citenamefont {Tanomura}, \citenamefont {Fukui}, \citenamefont {Ishimura}, \citenamefont {Sugiyama}, \citenamefont {Nakano},\ and\ \citenamefont {Tanemura}}]{Soma2022-hl}%
  \BibitemOpen
  \bibfield  {author} {\bibinfo {author} {\bibfnamefont {G.}~\bibnamefont {Soma}}, \bibinfo {author} {\bibfnamefont {W.}~\bibnamefont {Yanwachirakul}}, \bibinfo {author} {\bibfnamefont {T.}~\bibnamefont {Miyazaki}}, \bibinfo {author} {\bibfnamefont {E.}~\bibnamefont {Kato}}, \bibinfo {author} {\bibfnamefont {B.}~\bibnamefont {Onodera}}, \bibinfo {author} {\bibfnamefont {R.}~\bibnamefont {Tanomura}}, \bibinfo {author} {\bibfnamefont {T.}~\bibnamefont {Fukui}}, \bibinfo {author} {\bibfnamefont {S.}~\bibnamefont {Ishimura}}, \bibinfo {author} {\bibfnamefont {M.}~\bibnamefont {Sugiyama}}, \bibinfo {author} {\bibfnamefont {Y.}~\bibnamefont {Nakano}},\ and\ \bibinfo {author} {\bibfnamefont {T.}~\bibnamefont {Tanemura}},\ }\bibfield  {title} {\bibinfo {title} {Ultra-broadband surface-normal coherent optical receiver with nanometallic polarizers},\ }\href {https://doi.org/10.1021/acsphotonics.2c00759} {\bibfield  {journal} {\bibinfo  {journal} {ACS Photonics}\ }\textbf {\bibinfo {volume} {9}},\ \bibinfo {pages}
  {2842} (\bibinfo {year} {2022})}\BibitemShut {NoStop}%
\bibitem [{\citenamefont {Yun}\ \emph {et~al.}()\citenamefont {Yun}, \citenamefont {Roh}, \citenamefont {Lee}, \citenamefont {Park}, \citenamefont {Lim}, \citenamefont {Ahn},\ and\ \citenamefont {Choo}}]{Yun2021-mm}%
  \BibitemOpen
  \bibfield  {author} {\bibinfo {author} {\bibfnamefont {S.}~\bibnamefont {Yun}}, \bibinfo {author} {\bibfnamefont {S.}~\bibnamefont {Roh}}, \bibinfo {author} {\bibfnamefont {S.}~\bibnamefont {Lee}}, \bibinfo {author} {\bibfnamefont {H.}~\bibnamefont {Park}}, \bibinfo {author} {\bibfnamefont {M.}~\bibnamefont {Lim}}, \bibinfo {author} {\bibfnamefont {S.}~\bibnamefont {Ahn}},\ and\ \bibinfo {author} {\bibfnamefont {H.}~\bibnamefont {Choo}},\ }\bibfield  {title} {\bibinfo {title} {Highly efficient color separation and focusing in the sub-micron {CMOS} image sensor},\ }in\ \href {https://doi.org/10.1109/IEDM19574.2021.9720592} {\emph {\bibinfo {booktitle} {{IEEE} International Electron Devices Meeting ({IEDM}) 2021}}},\ pp.\ \bibinfo {pages} {30.1.1--30.1.4}\BibitemShut {NoStop}%
\bibitem [{\citenamefont {Uenoyama}\ and\ \citenamefont {Ota}(2022)}]{Uenoyama2022-fx}%
  \BibitemOpen
  \bibfield  {author} {\bibinfo {author} {\bibfnamefont {S.}~\bibnamefont {Uenoyama}}\ and\ \bibinfo {author} {\bibfnamefont {R.}~\bibnamefont {Ota}},\ }\bibfield  {title} {\bibinfo {title} {Monolithic integration of metalens in silicon photomultiplier for improved photodetection efficiency},\ }\href {https://doi.org/10.1002/adom.202102707} {\bibfield  {journal} {\bibinfo  {journal} {Adv. Opt. Mater.}\ }\textbf {\bibinfo {volume} {10}},\ \bibinfo {pages} {2102707} (\bibinfo {year} {2022})}\BibitemShut {NoStop}%
\bibitem [{\citenamefont {Yang}\ \emph {et~al.}(2023)\citenamefont {Yang}, \citenamefont {Seong}, \citenamefont {Choi}, \citenamefont {Park}, \citenamefont {Kim}, \citenamefont {Kim}, \citenamefont {Jeong}, \citenamefont {Jung}, \citenamefont {Kim}, \citenamefont {Jeon} \emph {et~al.}}]{yang2023integrated}%
  \BibitemOpen
  \bibfield  {author} {\bibinfo {author} {\bibfnamefont {Y.}~\bibnamefont {Yang}}, \bibinfo {author} {\bibfnamefont {J.}~\bibnamefont {Seong}}, \bibinfo {author} {\bibfnamefont {M.}~\bibnamefont {Choi}}, \bibinfo {author} {\bibfnamefont {J.}~\bibnamefont {Park}}, \bibinfo {author} {\bibfnamefont {G.}~\bibnamefont {Kim}}, \bibinfo {author} {\bibfnamefont {H.}~\bibnamefont {Kim}}, \bibinfo {author} {\bibfnamefont {J.}~\bibnamefont {Jeong}}, \bibinfo {author} {\bibfnamefont {C.}~\bibnamefont {Jung}}, \bibinfo {author} {\bibfnamefont {J.}~\bibnamefont {Kim}}, \bibinfo {author} {\bibfnamefont {G.}~\bibnamefont {Jeon}}, \emph {et~al.},\ }\bibfield  {title} {\bibinfo {title} {Integrated metasurfaces for re-envisioning a near-future disruptive optical platform},\ }\href {https://www.nature.com/articles/s41377-023-01169-4} {\bibfield  {journal} {\bibinfo  {journal} {Light Sci. Appl.}\ }\textbf {\bibinfo {volume} {12}},\ \bibinfo {pages} {152} (\bibinfo {year} {2023})}\BibitemShut {NoStop}%
\bibitem [{\citenamefont {Wenger}\ \emph {et~al.}(2022)\citenamefont {Wenger}, \citenamefont {Muller}, \citenamefont {Hill}, \citenamefont {Fisher}, \citenamefont {Ting}, \citenamefont {Wilson}, \citenamefont {Gunapala},\ and\ \citenamefont {Soibel}}]{Wenger2022-en}%
  \BibitemOpen
  \bibfield  {author} {\bibinfo {author} {\bibfnamefont {T.}~\bibnamefont {Wenger}}, \bibinfo {author} {\bibfnamefont {R.}~\bibnamefont {Muller}}, \bibinfo {author} {\bibfnamefont {C.~J.}\ \bibnamefont {Hill}}, \bibinfo {author} {\bibfnamefont {A.}~\bibnamefont {Fisher}}, \bibinfo {author} {\bibfnamefont {D.~Z.}\ \bibnamefont {Ting}}, \bibinfo {author} {\bibfnamefont {D.}~\bibnamefont {Wilson}}, \bibinfo {author} {\bibfnamefont {S.~D.}\ \bibnamefont {Gunapala}},\ and\ \bibinfo {author} {\bibfnamefont {A.}~\bibnamefont {Soibel}},\ }\bibfield  {title} {\bibinfo {title} {Infrared {nBn} detectors monolithically integrated with metasurface-based optical concentrators},\ }\href {https://doi.org/10.1063/5.0121643} {\bibfield  {journal} {\bibinfo  {journal} {Appl. Phys. Lett.}\ }\textbf {\bibinfo {volume} {121}},\ \bibinfo {pages} {181109} (\bibinfo {year} {2022})}\BibitemShut {NoStop}%
\bibitem [{\citenamefont {Lien}\ \emph {et~al.}(2024)\citenamefont {Lien}, \citenamefont {Wang}, \citenamefont {Wenger}, \citenamefont {Wang}, \citenamefont {Gunapala},\ and\ \citenamefont {Povinelli}}]{Lien2024-cd}%
  \BibitemOpen
  \bibfield  {author} {\bibinfo {author} {\bibfnamefont {M.~R.}\ \bibnamefont {Lien}}, \bibinfo {author} {\bibfnamefont {N.}~\bibnamefont {Wang}}, \bibinfo {author} {\bibfnamefont {T.}~\bibnamefont {Wenger}}, \bibinfo {author} {\bibfnamefont {H.}~\bibnamefont {Wang}}, \bibinfo {author} {\bibfnamefont {S.~D.}\ \bibnamefont {Gunapala}},\ and\ \bibinfo {author} {\bibfnamefont {M.~L.}\ \bibnamefont {Povinelli}},\ }\bibfield  {title} {\bibinfo {title} {An all‐silicon metalens integrated with a mid‐wave infrared black phosphorus photodiode},\ }\href {https://doi.org/10.1002/adom.202301952} {\bibfield  {journal} {\bibinfo  {journal} {Adv. Opt. Mater.}\ }\textbf {\bibinfo {volume} {12}},\ \bibinfo {pages} {2301952} (\bibinfo {year} {2024})}\BibitemShut {NoStop}%
\bibitem [{\citenamefont {Xie}\ \emph {et~al.}(2020)\citenamefont {Xie}, \citenamefont {Ni}, \citenamefont {Wang}, \citenamefont {Kan}, \citenamefont {Briere}, \citenamefont {Chen}, \citenamefont {Zhao}, \citenamefont {Delga}, \citenamefont {Ren}, \citenamefont {Chen}, \citenamefont {Xu},\ and\ \citenamefont {Genevet}}]{Xie2020-ga}%
  \BibitemOpen
  \bibfield  {author} {\bibinfo {author} {\bibfnamefont {Y.-Y.}\ \bibnamefont {Xie}}, \bibinfo {author} {\bibfnamefont {P.-N.}\ \bibnamefont {Ni}}, \bibinfo {author} {\bibfnamefont {Q.-H.}\ \bibnamefont {Wang}}, \bibinfo {author} {\bibfnamefont {Q.}~\bibnamefont {Kan}}, \bibinfo {author} {\bibfnamefont {G.}~\bibnamefont {Briere}}, \bibinfo {author} {\bibfnamefont {P.-P.}\ \bibnamefont {Chen}}, \bibinfo {author} {\bibfnamefont {Z.-Z.}\ \bibnamefont {Zhao}}, \bibinfo {author} {\bibfnamefont {A.}~\bibnamefont {Delga}}, \bibinfo {author} {\bibfnamefont {H.-R.}\ \bibnamefont {Ren}}, \bibinfo {author} {\bibfnamefont {H.-D.}\ \bibnamefont {Chen}}, \bibinfo {author} {\bibfnamefont {C.}~\bibnamefont {Xu}},\ and\ \bibinfo {author} {\bibfnamefont {P.}~\bibnamefont {Genevet}},\ }\bibfield  {title} {\bibinfo {title} {Metasurface-integrated vertical cavity surface-emitting lasers for programmable directional lasing emissions},\ }\href {https://doi.org/10.1038/s41565-019-0611-y} {\bibfield  {journal} {\bibinfo  {journal}
  {Nat. Nanotechnol.}\ }\textbf {\bibinfo {volume} {15}},\ \bibinfo {pages} {125} (\bibinfo {year} {2020})}\BibitemShut {NoStop}%
\bibitem [{\citenamefont {Wen}\ \emph {et~al.}(2021)\citenamefont {Wen}, \citenamefont {Meng}, \citenamefont {Cadusch},\ and\ \citenamefont {Crozier}}]{Wen2021-uk}%
  \BibitemOpen
  \bibfield  {author} {\bibinfo {author} {\bibfnamefont {D.}~\bibnamefont {Wen}}, \bibinfo {author} {\bibfnamefont {J.}~\bibnamefont {Meng}}, \bibinfo {author} {\bibfnamefont {J.~J.}\ \bibnamefont {Cadusch}},\ and\ \bibinfo {author} {\bibfnamefont {K.~B.}\ \bibnamefont {Crozier}},\ }\bibfield  {title} {\bibinfo {title} {{VCSELs} with on‐facet metasurfaces for polarization state generation and detection},\ }\href {https://doi.org/10.1002/adom.202001780} {\bibfield  {journal} {\bibinfo  {journal} {Adv. Opt. Mater.}\ }\textbf {\bibinfo {volume} {9}},\ \bibinfo {pages} {2001780} (\bibinfo {year} {2021})}\BibitemShut {NoStop}%
\bibitem [{\citenamefont {Fu}\ \emph {et~al.}(2023)\citenamefont {Fu}, \citenamefont {Ni}, \citenamefont {Wu}, \citenamefont {Pei}, \citenamefont {Wang}, \citenamefont {Chen}, \citenamefont {Xu}, \citenamefont {Kan}, \citenamefont {Chu},\ and\ \citenamefont {Xie}}]{fu2023metasurface}%
  \BibitemOpen
  \bibfield  {author} {\bibinfo {author} {\bibfnamefont {P.}~\bibnamefont {Fu}}, \bibinfo {author} {\bibfnamefont {P.-N.}\ \bibnamefont {Ni}}, \bibinfo {author} {\bibfnamefont {B.}~\bibnamefont {Wu}}, \bibinfo {author} {\bibfnamefont {X.-Z.}\ \bibnamefont {Pei}}, \bibinfo {author} {\bibfnamefont {Q.-H.}\ \bibnamefont {Wang}}, \bibinfo {author} {\bibfnamefont {P.-P.}\ \bibnamefont {Chen}}, \bibinfo {author} {\bibfnamefont {C.}~\bibnamefont {Xu}}, \bibinfo {author} {\bibfnamefont {Q.}~\bibnamefont {Kan}}, \bibinfo {author} {\bibfnamefont {W.-G.}\ \bibnamefont {Chu}},\ and\ \bibinfo {author} {\bibfnamefont {Y.-Y.}\ \bibnamefont {Xie}},\ }\bibfield  {title} {\bibinfo {title} {Metasurface enabled on-chip generation and manipulation of vector beams from vertical cavity surface-emitting lasers},\ }\href {https://advanced.onlinelibrary.wiley.com/doi/abs/10.1002/adma.202204286} {\bibfield  {journal} {\bibinfo  {journal} {Adv. Mater.}\ }\textbf {\bibinfo {volume} {35}},\ \bibinfo {pages} {2204286} (\bibinfo {year}
  {2023})}\BibitemShut {NoStop}%
\bibitem [{\citenamefont {Zheng}\ \emph {et~al.}(2025)\citenamefont {Zheng}, \citenamefont {Ni}, \citenamefont {Xie},\ and\ \citenamefont {Genevet}}]{zheng2025chip}%
  \BibitemOpen
  \bibfield  {author} {\bibinfo {author} {\bibfnamefont {C.-L.}\ \bibnamefont {Zheng}}, \bibinfo {author} {\bibfnamefont {P.-N.}\ \bibnamefont {Ni}}, \bibinfo {author} {\bibfnamefont {Y.-Y.}\ \bibnamefont {Xie}},\ and\ \bibinfo {author} {\bibfnamefont {P.}~\bibnamefont {Genevet}},\ }\bibfield  {title} {\bibinfo {title} {On-chip light control of semiconductor optoelectronic devices using integrated metasurfaces},\ }\href {http://oejournal.org//article/doi/10.29026/oea.2025.240159} {\bibfield  {journal} {\bibinfo  {journal} {Opt. Elect. Adv.}\ ,\ \bibinfo {pages} {240159}} (\bibinfo {year} {2025})}\BibitemShut {NoStop}%
\bibitem [{\citenamefont {Winzer}\ and\ \citenamefont {Neilson}(2017)}]{Winzer2017-tx}%
  \BibitemOpen
  \bibfield  {author} {\bibinfo {author} {\bibfnamefont {P.~J.}\ \bibnamefont {Winzer}}\ and\ \bibinfo {author} {\bibfnamefont {D.~T.}\ \bibnamefont {Neilson}},\ }\bibfield  {title} {\bibinfo {title} {From scaling disparities to integrated parallelism: {A} decathlon for a decade},\ }\href {https://doi.org/10.1109/jlt.2017.2662082} {\bibfield  {journal} {\bibinfo  {journal} {J. Lightwave Technol.}\ }\textbf {\bibinfo {volume} {35}},\ \bibinfo {pages} {1099} (\bibinfo {year} {2017})}\BibitemShut {NoStop}%
\bibitem [{\citenamefont {Miller}(2017)}]{Miller2017-gy}%
  \BibitemOpen
  \bibfield  {author} {\bibinfo {author} {\bibfnamefont {D.~A.~B.}\ \bibnamefont {Miller}},\ }\bibfield  {title} {\bibinfo {title} {Attojoule optoelectronics for low-energy information processing and communications},\ }\href {https://doi.org/10.1109/jlt.2017.2647779} {\bibfield  {journal} {\bibinfo  {journal} {J. Lightwave Technol.}\ }\textbf {\bibinfo {volume} {35}},\ \bibinfo {pages} {346} (\bibinfo {year} {2017})}\BibitemShut {NoStop}%
\bibitem [{\citenamefont {Winzer}(2023)}]{Winzer2023-qf}%
  \BibitemOpen
  \bibfield  {author} {\bibinfo {author} {\bibfnamefont {P.~J.}\ \bibnamefont {Winzer}},\ }\bibfield  {title} {\bibinfo {title} {The future of communications is massively parallel},\ }\href {https://doi.org/10.1364/jocn.496992} {\bibfield  {journal} {\bibinfo  {journal} {J. Opt. Commun. Netw.}\ }\textbf {\bibinfo {volume} {15}},\ \bibinfo {pages} {783} (\bibinfo {year} {2023})}\BibitemShut {NoStop}%
\bibitem [{\citenamefont {Wang}\ \emph {et~al.}(2012)\citenamefont {Wang}, \citenamefont {Yang}, \citenamefont {Fazal}, \citenamefont {Ahmed}, \citenamefont {Yan}, \citenamefont {Huang}, \citenamefont {Ren}, \citenamefont {Yue}, \citenamefont {Dolinar}, \citenamefont {Tur},\ and\ \citenamefont {Willner}}]{Wang2012-py}%
  \BibitemOpen
  \bibfield  {author} {\bibinfo {author} {\bibfnamefont {J.}~\bibnamefont {Wang}}, \bibinfo {author} {\bibfnamefont {J.-Y.}\ \bibnamefont {Yang}}, \bibinfo {author} {\bibfnamefont {I.~M.}\ \bibnamefont {Fazal}}, \bibinfo {author} {\bibfnamefont {N.}~\bibnamefont {Ahmed}}, \bibinfo {author} {\bibfnamefont {Y.}~\bibnamefont {Yan}}, \bibinfo {author} {\bibfnamefont {H.}~\bibnamefont {Huang}}, \bibinfo {author} {\bibfnamefont {Y.}~\bibnamefont {Ren}}, \bibinfo {author} {\bibfnamefont {Y.}~\bibnamefont {Yue}}, \bibinfo {author} {\bibfnamefont {S.}~\bibnamefont {Dolinar}}, \bibinfo {author} {\bibfnamefont {M.}~\bibnamefont {Tur}},\ and\ \bibinfo {author} {\bibfnamefont {A.~E.}\ \bibnamefont {Willner}},\ }\bibfield  {title} {\bibinfo {title} {Terabit free-space data transmission employing orbital angular momentum multiplexing},\ }\href {https://doi.org/10.1038/nphoton.2012.138} {\bibfield  {journal} {\bibinfo  {journal} {Nat. Photonics}\ }\textbf {\bibinfo {volume} {6}},\ \bibinfo {pages} {488} (\bibinfo {year}
  {2012})}\BibitemShut {NoStop}%
\bibitem [{\citenamefont {He}\ \emph {et~al.}(2020)\citenamefont {He}, \citenamefont {Wang}, \citenamefont {Wang}, \citenamefont {Liu}, \citenamefont {Ye}, \citenamefont {Zhou}, \citenamefont {Li}, \citenamefont {Chen}, \citenamefont {Zhang},\ and\ \citenamefont {Fan}}]{He2020-wc}%
  \BibitemOpen
  \bibfield  {author} {\bibinfo {author} {\bibfnamefont {Y.}~\bibnamefont {He}}, \bibinfo {author} {\bibfnamefont {P.}~\bibnamefont {Wang}}, \bibinfo {author} {\bibfnamefont {C.}~\bibnamefont {Wang}}, \bibinfo {author} {\bibfnamefont {J.}~\bibnamefont {Liu}}, \bibinfo {author} {\bibfnamefont {H.}~\bibnamefont {Ye}}, \bibinfo {author} {\bibfnamefont {X.}~\bibnamefont {Zhou}}, \bibinfo {author} {\bibfnamefont {Y.}~\bibnamefont {Li}}, \bibinfo {author} {\bibfnamefont {S.}~\bibnamefont {Chen}}, \bibinfo {author} {\bibfnamefont {X.}~\bibnamefont {Zhang}},\ and\ \bibinfo {author} {\bibfnamefont {D.}~\bibnamefont {Fan}},\ }\bibfield  {title} {\bibinfo {title} {All-optical signal processing in structured light multiplexing with dielectric meta-optics},\ }\href {https://doi.org/10.1021/acsphotonics.9b01292} {\bibfield  {journal} {\bibinfo  {journal} {ACS Photonics}\ }\textbf {\bibinfo {volume} {7}},\ \bibinfo {pages} {135} (\bibinfo {year} {2020})}\BibitemShut {NoStop}%
\bibitem [{\citenamefont {Chen}\ \emph {et~al.}(2021)\citenamefont {Chen}, \citenamefont {Xie}, \citenamefont {Ye}, \citenamefont {Wang}, \citenamefont {Guo}, \citenamefont {He}, \citenamefont {Li}, \citenamefont {Yuan},\ and\ \citenamefont {Fan}}]{Chen2021-sa}%
  \BibitemOpen
  \bibfield  {author} {\bibinfo {author} {\bibfnamefont {S.}~\bibnamefont {Chen}}, \bibinfo {author} {\bibfnamefont {Z.}~\bibnamefont {Xie}}, \bibinfo {author} {\bibfnamefont {H.}~\bibnamefont {Ye}}, \bibinfo {author} {\bibfnamefont {X.}~\bibnamefont {Wang}}, \bibinfo {author} {\bibfnamefont {Z.}~\bibnamefont {Guo}}, \bibinfo {author} {\bibfnamefont {Y.}~\bibnamefont {He}}, \bibinfo {author} {\bibfnamefont {Y.}~\bibnamefont {Li}}, \bibinfo {author} {\bibfnamefont {X.}~\bibnamefont {Yuan}},\ and\ \bibinfo {author} {\bibfnamefont {D.}~\bibnamefont {Fan}},\ }\bibfield  {title} {\bibinfo {title} {Cylindrical vector beam multiplexer/demultiplexer using off-axis polarization control},\ }\href {https://doi.org/10.1038/s41377-021-00667-7} {\bibfield  {journal} {\bibinfo  {journal} {Light Sci Appl}\ }\textbf {\bibinfo {volume} {10}},\ \bibinfo {pages} {222} (\bibinfo {year} {2021})}\BibitemShut {NoStop}%
\bibitem [{\citenamefont {Horst}\ \emph {et~al.}(2023)\citenamefont {Horst}, \citenamefont {Bitachon}, \citenamefont {Kulmer}, \citenamefont {Brun}, \citenamefont {Blatter}, \citenamefont {Conan}, \citenamefont {Montmerle-Bonnefois}, \citenamefont {Montri}, \citenamefont {Sorrente}, \citenamefont {Lim}, \citenamefont {Védrenne}, \citenamefont {Matter}, \citenamefont {Pommarel}, \citenamefont {Baeuerle},\ and\ \citenamefont {Leuthold}}]{Horst2023-dp}%
  \BibitemOpen
  \bibfield  {author} {\bibinfo {author} {\bibfnamefont {Y.}~\bibnamefont {Horst}}, \bibinfo {author} {\bibfnamefont {B.~I.}\ \bibnamefont {Bitachon}}, \bibinfo {author} {\bibfnamefont {L.}~\bibnamefont {Kulmer}}, \bibinfo {author} {\bibfnamefont {J.}~\bibnamefont {Brun}}, \bibinfo {author} {\bibfnamefont {T.}~\bibnamefont {Blatter}}, \bibinfo {author} {\bibfnamefont {J.-M.}\ \bibnamefont {Conan}}, \bibinfo {author} {\bibfnamefont {A.}~\bibnamefont {Montmerle-Bonnefois}}, \bibinfo {author} {\bibfnamefont {J.}~\bibnamefont {Montri}}, \bibinfo {author} {\bibfnamefont {B.}~\bibnamefont {Sorrente}}, \bibinfo {author} {\bibfnamefont {C.~B.}\ \bibnamefont {Lim}}, \bibinfo {author} {\bibfnamefont {N.}~\bibnamefont {Védrenne}}, \bibinfo {author} {\bibfnamefont {D.}~\bibnamefont {Matter}}, \bibinfo {author} {\bibfnamefont {L.}~\bibnamefont {Pommarel}}, \bibinfo {author} {\bibfnamefont {B.}~\bibnamefont {Baeuerle}},\ and\ \bibinfo {author} {\bibfnamefont {J.}~\bibnamefont {Leuthold}},\ }\bibfield  {title} {\bibinfo
  {title} {{Tbit/s} line-rate satellite feeder links enabled by coherent modulation and full-adaptive optics},\ }\href {https://doi.org/10.1038/s41377-023-01201-7} {\bibfield  {journal} {\bibinfo  {journal} {Light Sci. Appl.}\ }\textbf {\bibinfo {volume} {12}},\ \bibinfo {pages} {153} (\bibinfo {year} {2023})}\BibitemShut {NoStop}%
\bibitem [{\citenamefont {Hamerly}\ \emph {et~al.}(2019)\citenamefont {Hamerly}, \citenamefont {Bernstein}, \citenamefont {Sludds}, \citenamefont {Solja{\v c}i{\'c}},\ and\ \citenamefont {Englund}}]{Hamerly2019-vs}%
  \BibitemOpen
  \bibfield  {author} {\bibinfo {author} {\bibfnamefont {R.}~\bibnamefont {Hamerly}}, \bibinfo {author} {\bibfnamefont {L.}~\bibnamefont {Bernstein}}, \bibinfo {author} {\bibfnamefont {A.}~\bibnamefont {Sludds}}, \bibinfo {author} {\bibfnamefont {M.}~\bibnamefont {Solja{\v c}i{\'c}}},\ and\ \bibinfo {author} {\bibfnamefont {D.}~\bibnamefont {Englund}},\ }\bibfield  {title} {\bibinfo {title} {Large-scale optical neural networks based on photoelectric multiplication},\ }\href {https://doi.org/10.1103/PhysRevX.9.021032} {\bibfield  {journal} {\bibinfo  {journal} {Phys. Rev. X}\ }\textbf {\bibinfo {volume} {9}},\ \bibinfo {pages} {021032} (\bibinfo {year} {2019})}\BibitemShut {NoStop}%
\bibitem [{\citenamefont {Wang}\ \emph {et~al.}(2022)\citenamefont {Wang}, \citenamefont {Ma}, \citenamefont {Wright}, \citenamefont {Onodera}, \citenamefont {Richard},\ and\ \citenamefont {McMahon}}]{Wang2022-qx}%
  \BibitemOpen
  \bibfield  {author} {\bibinfo {author} {\bibfnamefont {T.}~\bibnamefont {Wang}}, \bibinfo {author} {\bibfnamefont {S.-Y.}\ \bibnamefont {Ma}}, \bibinfo {author} {\bibfnamefont {L.~G.}\ \bibnamefont {Wright}}, \bibinfo {author} {\bibfnamefont {T.}~\bibnamefont {Onodera}}, \bibinfo {author} {\bibfnamefont {B.~C.}\ \bibnamefont {Richard}},\ and\ \bibinfo {author} {\bibfnamefont {P.~L.}\ \bibnamefont {McMahon}},\ }\bibfield  {title} {\bibinfo {title} {An optical neural network using less than 1 photon per multiplication},\ }\href {https://doi.org/10.1038/s41467-021-27774-8} {\bibfield  {journal} {\bibinfo  {journal} {Nat. Commun.}\ }\textbf {\bibinfo {volume} {13}},\ \bibinfo {pages} {123} (\bibinfo {year} {2022})}\BibitemShut {NoStop}%
\bibitem [{\citenamefont {Chen}\ \emph {et~al.}(2023)\citenamefont {Chen}, \citenamefont {Sludds}, \citenamefont {Davis}, \citenamefont {Christen}, \citenamefont {Bernstein}, \citenamefont {Ateshian}, \citenamefont {Heuser}, \citenamefont {Heermeier}, \citenamefont {Lott}, \citenamefont {Reitzenstein}, \citenamefont {Hamerly},\ and\ \citenamefont {Englund}}]{Chen2023-lk}%
  \BibitemOpen
  \bibfield  {author} {\bibinfo {author} {\bibfnamefont {Z.}~\bibnamefont {Chen}}, \bibinfo {author} {\bibfnamefont {A.}~\bibnamefont {Sludds}}, \bibinfo {author} {\bibfnamefont {R.}~\bibnamefont {Davis}}, \bibinfo {author} {\bibfnamefont {I.}~\bibnamefont {Christen}}, \bibinfo {author} {\bibfnamefont {L.}~\bibnamefont {Bernstein}}, \bibinfo {author} {\bibfnamefont {L.}~\bibnamefont {Ateshian}}, \bibinfo {author} {\bibfnamefont {T.}~\bibnamefont {Heuser}}, \bibinfo {author} {\bibfnamefont {N.}~\bibnamefont {Heermeier}}, \bibinfo {author} {\bibfnamefont {J.~A.}\ \bibnamefont {Lott}}, \bibinfo {author} {\bibfnamefont {S.}~\bibnamefont {Reitzenstein}}, \bibinfo {author} {\bibfnamefont {R.}~\bibnamefont {Hamerly}},\ and\ \bibinfo {author} {\bibfnamefont {D.}~\bibnamefont {Englund}},\ }\bibfield  {title} {\bibinfo {title} {Deep learning with coherent {VCSEL} neural networks},\ }\href {https://doi.org/10.1038/s41566-023-01233-w} {\bibfield  {journal} {\bibinfo  {journal} {Nat. Photonics}\ }\textbf {\bibinfo
  {volume} {17}},\ \bibinfo {pages} {723} (\bibinfo {year} {2023})}\BibitemShut {NoStop}%
\bibitem [{\citenamefont {Martin}\ \emph {et~al.}(2018)\citenamefont {Martin}, \citenamefont {Dodane}, \citenamefont {Leviandier}, \citenamefont {Dolfi}, \citenamefont {Naughton}, \citenamefont {O'Brien}, \citenamefont {Spuessens}, \citenamefont {Baets}, \citenamefont {Lepage}, \citenamefont {Verheyen}, \citenamefont {De~Heyn}, \citenamefont {Absil}, \citenamefont {Feneyrou},\ and\ \citenamefont {Bourderionnet}}]{Martin2018-jx}%
  \BibitemOpen
  \bibfield  {author} {\bibinfo {author} {\bibfnamefont {A.}~\bibnamefont {Martin}}, \bibinfo {author} {\bibfnamefont {D.}~\bibnamefont {Dodane}}, \bibinfo {author} {\bibfnamefont {L.}~\bibnamefont {Leviandier}}, \bibinfo {author} {\bibfnamefont {D.}~\bibnamefont {Dolfi}}, \bibinfo {author} {\bibfnamefont {A.}~\bibnamefont {Naughton}}, \bibinfo {author} {\bibfnamefont {P.}~\bibnamefont {O'Brien}}, \bibinfo {author} {\bibfnamefont {T.}~\bibnamefont {Spuessens}}, \bibinfo {author} {\bibfnamefont {R.}~\bibnamefont {Baets}}, \bibinfo {author} {\bibfnamefont {G.}~\bibnamefont {Lepage}}, \bibinfo {author} {\bibfnamefont {P.}~\bibnamefont {Verheyen}}, \bibinfo {author} {\bibfnamefont {P.}~\bibnamefont {De~Heyn}}, \bibinfo {author} {\bibfnamefont {P.}~\bibnamefont {Absil}}, \bibinfo {author} {\bibfnamefont {P.}~\bibnamefont {Feneyrou}},\ and\ \bibinfo {author} {\bibfnamefont {J.}~\bibnamefont {Bourderionnet}},\ }\bibfield  {title} {\bibinfo {title} {Photonic integrated circuit-based {FMCW} coherent {LiDAR}},\ }\href
  {https://doi.org/10.1109/JLT.2018.2840223} {\bibfield  {journal} {\bibinfo  {journal} {J. Lightwave Technol.}\ }\textbf {\bibinfo {volume} {36}},\ \bibinfo {pages} {4640} (\bibinfo {year} {2018})}\BibitemShut {NoStop}%
\bibitem [{\citenamefont {Riemensberger}\ \emph {et~al.}(2020)\citenamefont {Riemensberger}, \citenamefont {Lukashchuk}, \citenamefont {Karpov}, \citenamefont {Weng}, \citenamefont {Lucas}, \citenamefont {Liu},\ and\ \citenamefont {Kippenberg}}]{Riemensberger2020-vi}%
  \BibitemOpen
  \bibfield  {author} {\bibinfo {author} {\bibfnamefont {J.}~\bibnamefont {Riemensberger}}, \bibinfo {author} {\bibfnamefont {A.}~\bibnamefont {Lukashchuk}}, \bibinfo {author} {\bibfnamefont {M.}~\bibnamefont {Karpov}}, \bibinfo {author} {\bibfnamefont {W.}~\bibnamefont {Weng}}, \bibinfo {author} {\bibfnamefont {E.}~\bibnamefont {Lucas}}, \bibinfo {author} {\bibfnamefont {J.}~\bibnamefont {Liu}},\ and\ \bibinfo {author} {\bibfnamefont {T.~J.}\ \bibnamefont {Kippenberg}},\ }\bibfield  {title} {\bibinfo {title} {Massively parallel coherent laser ranging using a soliton microcomb},\ }\href {https://doi.org/10.1038/s41586-020-2239-3} {\bibfield  {journal} {\bibinfo  {journal} {Nature}\ }\textbf {\bibinfo {volume} {581}},\ \bibinfo {pages} {164} (\bibinfo {year} {2020})}\BibitemShut {NoStop}%
\bibitem [{\citenamefont {Rogers}\ \emph {et~al.}(2021)\citenamefont {Rogers}, \citenamefont {Piggott}, \citenamefont {Thomson}, \citenamefont {Wiser}, \citenamefont {Opris}, \citenamefont {Fortune}, \citenamefont {Compston}, \citenamefont {Gondarenko}, \citenamefont {Meng}, \citenamefont {Chen}, \citenamefont {Reed},\ and\ \citenamefont {Nicolaescu}}]{Rogers2021-gk}%
  \BibitemOpen
  \bibfield  {author} {\bibinfo {author} {\bibfnamefont {C.}~\bibnamefont {Rogers}}, \bibinfo {author} {\bibfnamefont {A.~Y.}\ \bibnamefont {Piggott}}, \bibinfo {author} {\bibfnamefont {D.~J.}\ \bibnamefont {Thomson}}, \bibinfo {author} {\bibfnamefont {R.~F.}\ \bibnamefont {Wiser}}, \bibinfo {author} {\bibfnamefont {I.~E.}\ \bibnamefont {Opris}}, \bibinfo {author} {\bibfnamefont {S.~A.}\ \bibnamefont {Fortune}}, \bibinfo {author} {\bibfnamefont {A.~J.}\ \bibnamefont {Compston}}, \bibinfo {author} {\bibfnamefont {A.}~\bibnamefont {Gondarenko}}, \bibinfo {author} {\bibfnamefont {F.}~\bibnamefont {Meng}}, \bibinfo {author} {\bibfnamefont {X.}~\bibnamefont {Chen}}, \bibinfo {author} {\bibfnamefont {G.~T.}\ \bibnamefont {Reed}},\ and\ \bibinfo {author} {\bibfnamefont {R.}~\bibnamefont {Nicolaescu}},\ }\bibfield  {title} {\bibinfo {title} {A universal {3D} imaging sensor on a silicon photonics platform},\ }\href {https://doi.org/10.1038/s41586-021-03259-y} {\bibfield  {journal} {\bibinfo  {journal} {Nature}\
  }\textbf {\bibinfo {volume} {590}},\ \bibinfo {pages} {256} (\bibinfo {year} {2021})}\BibitemShut {NoStop}%
\bibitem [{\citenamefont {Han}\ \emph {et~al.}(2017)\citenamefont {Han}, \citenamefont {Boeuf}, \citenamefont {Fujikata}, \citenamefont {Takahashi}, \citenamefont {Takagi},\ and\ \citenamefont {Takenaka}}]{Han2017-pt}%
  \BibitemOpen
  \bibfield  {author} {\bibinfo {author} {\bibfnamefont {J.-H.}\ \bibnamefont {Han}}, \bibinfo {author} {\bibfnamefont {F.}~\bibnamefont {Boeuf}}, \bibinfo {author} {\bibfnamefont {J.}~\bibnamefont {Fujikata}}, \bibinfo {author} {\bibfnamefont {S.}~\bibnamefont {Takahashi}}, \bibinfo {author} {\bibfnamefont {S.}~\bibnamefont {Takagi}},\ and\ \bibinfo {author} {\bibfnamefont {M.}~\bibnamefont {Takenaka}},\ }\bibfield  {title} {\bibinfo {title} {Efficient low-loss {InGaAsP/Si} hybrid {MOS} optical modulator},\ }\href {https://doi.org/10.1038/nphoton.2017.122} {\bibfield  {journal} {\bibinfo  {journal} {Nat. Photonics}\ }\textbf {\bibinfo {volume} {11}},\ \bibinfo {pages} {486} (\bibinfo {year} {2017})}\BibitemShut {NoStop}%
\bibitem [{\citenamefont {Ravi}\ \emph {et~al.}(2004)\citenamefont {Ravi}, \citenamefont {DasGupta},\ and\ \citenamefont {DasGupta}}]{Ravi2004-le}%
  \BibitemOpen
  \bibfield  {author} {\bibinfo {author} {\bibfnamefont {M.~R.}\ \bibnamefont {Ravi}}, \bibinfo {author} {\bibfnamefont {A.}~\bibnamefont {DasGupta}},\ and\ \bibinfo {author} {\bibfnamefont {N.}~\bibnamefont {DasGupta}},\ }\bibfield  {title} {\bibinfo {title} {Silicon nitride and polyimide capping layers on {InGaAs/InP} {PIN} photodetector after sulfur treatment},\ }\href {https://doi.org/10.1016/j.jcrysgro.2004.04.054} {\bibfield  {journal} {\bibinfo  {journal} {J. Cryst. Growth}\ }\textbf {\bibinfo {volume} {268}},\ \bibinfo {pages} {359} (\bibinfo {year} {2004})}\BibitemShut {NoStop}%
\bibitem [{\citenamefont {Liu}\ and\ \citenamefont {Fan}(2012)}]{Liu2012-kd}%
  \BibitemOpen
  \bibfield  {author} {\bibinfo {author} {\bibfnamefont {V.}~\bibnamefont {Liu}}\ and\ \bibinfo {author} {\bibfnamefont {S.}~\bibnamefont {Fan}},\ }\bibfield  {title} {\bibinfo {title} {S$^4$: {A} free electromagnetic solver for layered periodic structures},\ }\href {https://doi.org/10.1016/j.cpc.2012.04.026} {\bibfield  {journal} {\bibinfo  {journal} {Comput. Phys. Commun.}\ }\textbf {\bibinfo {volume} {183}},\ \bibinfo {pages} {2233} (\bibinfo {year} {2012})}\BibitemShut {NoStop}%
\bibitem [{\citenamefont {Mori}\ \emph {et~al.}(2012)\citenamefont {Mori}, \citenamefont {Zhang},\ and\ \citenamefont {Kikuchi}}]{Mori2012-fn}%
  \BibitemOpen
  \bibfield  {author} {\bibinfo {author} {\bibfnamefont {Y.}~\bibnamefont {Mori}}, \bibinfo {author} {\bibfnamefont {C.}~\bibnamefont {Zhang}},\ and\ \bibinfo {author} {\bibfnamefont {K.}~\bibnamefont {Kikuchi}},\ }\bibfield  {title} {\bibinfo {title} {Novel configuration of finite-impulse-response filters tolerant to carrier-phase fluctuations in digital coherent optical receivers for higher-order quadrature amplitude modulation signals},\ }\href {https://doi.org/10.1364/OE.20.026236} {\bibfield  {journal} {\bibinfo  {journal} {Opt. Express}\ }\textbf {\bibinfo {volume} {20}},\ \bibinfo {pages} {26236} (\bibinfo {year} {2012})}\BibitemShut {NoStop}%
\bibitem [{\citenamefont {Faruk}\ and\ \citenamefont {Kikuchi}(2013)}]{Faruk2013-uu}%
  \BibitemOpen
  \bibfield  {author} {\bibinfo {author} {\bibfnamefont {M.~S.}\ \bibnamefont {Faruk}}\ and\ \bibinfo {author} {\bibfnamefont {K.}~\bibnamefont {Kikuchi}},\ }\bibfield  {title} {\bibinfo {title} {Compensation for in-phase/quadrature imbalance in coherent-receiver front end for optical quadrature amplitude modulation},\ }\href {https://doi.org/10.1109/JPHOT.2013.2251872} {\bibfield  {journal} {\bibinfo  {journal} {IEEE Photonics J.}\ }\textbf {\bibinfo {volume} {5}},\ \bibinfo {pages} {7800110} (\bibinfo {year} {2013})}\BibitemShut {NoStop}%
\end{thebibliography}%

\clearpage

\setcounter{figure}{0}
\setcounter{table}{0}
\renewcommand{\figurename}{Extended Data Fig.}
\renewcommand{\tablename}{Extended Data Table}

\section*{Methods}

\subsection*{Device fabrication}
\begin{figure}[tb]
\centering\includegraphics{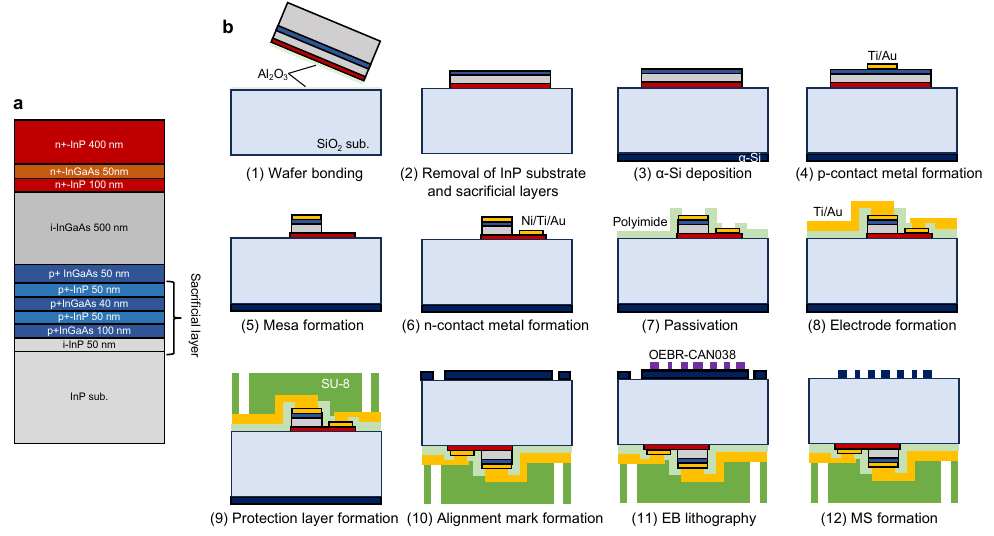}
\caption{
\textbf{Device fabrication.} 
\figlabel{a} Epitaxial layers grown on an InP substrate. The InP substrate and the sacrificial InP/InGaAs layers were wet-etched after the wafer bonding process.
\figlabel{b} Fabrication flow chart.
}
\label{fig:fabrication}
\end{figure}

\noindent
First, the epitaxial InGaAs/InP layers, consisting of p-i-n diode and sacrificial layers, were grown on a 2-inch InP substrate by metal-organic vapor phase epitaxy (MOVPE). The detailed layer structure is provided in Extended Data Fig.~\ref{fig:fabrication}a.
The device was then fabricated through the following steps as shown in Extended Data Fig.~\ref{fig:fabrication}b.

\noindent
\begin{enumerate}
\item {\it Wafer bonding}: The 2-inch InP substrate with p-i-n epitaxial layers was bonded on a 3-inch fused silica (SiO$_2$) substrate with an Al$_2$O$_3$ bonding interface \cite{Han2017-pt}.
\item {\it Removal of InP substrate and sacrificial layers}: The InP substrate and the sacrificial InP/InGaAs layers were removed through wet chemical etching using HCl (for InP) and H$_3$PO$_4$+H$_2$O$_2$ (for InGaAs) solutions. After this process, thin InGaAs/InP layers for the p-i-n membrane PDs remained on the SiO$_2$ substrate.
\item {\it $\alpha$-Si deposition}: A 1050-nm thick $\mathrm{\alpha}$-Si layer was deposited by plasma-enhanced chemical vapor deposition (PECVD) on the backside of the SiO$_2$ substrate, which is then protected with a polyimide layer.
The bonded wafer was diced into 1.2~cm squared chips for subsequent device processes.
\item {\it p-contact metal formation}: Ti/Au (30/150~nm) layers for contacting the p$^+$-InGaAs layer were formed by photolithography, electron beam (EB) evaporation, and liftoff process, followed by annealing at 300~$^\circ$C for 1~min to reduce the contact resistance.
\item {\it Mesa formation}: The PD mesas were formed through wet chemical etching with HCl~+~H$_3$PO$_4$ (for InP) and H$_2$SO$_4$~+~H$_2$O$_2$ (for InGaAs). 
Then, the larger mesas for n-contact were formed by photolithography and wet chemical etching.
\item {\it n-contact metal formation}: Ni/Ti/Au (30/20/150 nm) layers for contacting the n$^+$-InGaAs layer were formed in the same manner as the p-contact metal formation and annealed at 300~$^\circ$C for 1~min.
\item {\it Passivation}: After the native oxides on the surface were removed by buffered hydrofluoric acid (BHF) for 1~min, the surface was passivated through (NH$_4$)$_2$S$_x$ treatment for 30~min at room temperature \cite{Ravi2004-le}. 
Then, a photosensitive polyimide layer (LT-S8010A, Toray) was coated over the device. After forming the contact openings through photolithography, the device was cured at 230~$^\circ$C for 1~hour under a nitrogen atmosphere.
\item {\it Electrode formation}: Ti/Au (50/450~nm) electrode patterns were formed by photolithography, EB evaporation, and liftoff process.
\item {\it Protection layer formation}: A thick photoresist layer (SU-8~3005, Kayaku Advanced Materials) was spin-coated on the device as a protection layer. The contact openings at the electrode pads were formed through photolithography.
\item {\it Alignment mark formation}: The chip was flipped and the backside was cleaned through the O$_2$ ashing process.
The alignment marks were then formed through photolithography with backside alignment to the PD patterns, followed by reactive-ion etching (RIE) of the $\mathrm{\alpha}$-Si layer.
\item {\it EB lithography}:  
The MS patterns were defined by EB lithography (F7000S, ADVANTEST) using negative EB resist (OBER-CAN038, Tokyo Ohka Kogyo) and developer (NMD-W).
\item {\it MS formation}: 
The MS patterns on EB resist were transferred to the $\mathrm{\alpha}$-Si layer by RIE (MUC-21 ASE-SRE, Sumitomo Precision Products) with SF$_6$ and C$_4$F$_8$ gases, known as the Bosch process, followed by the O$_2$ ashing process.
\end{enumerate}

\subsection*{MS design}
\subsubsection*{ML}
\noindent
To design an ML that focuses incident lightwave to a PD at a desired position, we first calculate the phase shift required at each position $(x,y)$ of the MS. It is described as the sum of two phase profiles: $\varphi_\mathrm{ML}(x,y)=\varphi_\mathrm{col}(x,y)+\varphi_\mathrm{foc}(x,y)$. The first term $\varphi_\mathrm{col}(x,y)$ corresponds to the function of collimating a spherical lightwave from a fiber core located at a distance $f_1$ and is expressed as
\begin{align}
    \varphi_\mathrm{col}(x,y) = -\frac{2\pi}{\lambda}\left(\sqrt{x^2+y^2+f_1^2}-f_1\right),
    \label{eq:phase0}
\end{align}
where $\lambda$ is the wavelength.
On the other hand, the second term $\varphi_\mathrm{foc}(x,y)$ corresponds to the spherical phase profile to achieve the function of a focusing lens with a focal length of $f_2$ inside the SiO$_2$ substrate. Using $n_s$ to denote the refractive index of SiO$_2$, $\varphi_\mathrm{col}(x,y)$ is expressed as
\begin{align}
    \varphi_\mathrm{foc}(x,y) = -\frac{2n_s\pi}{\lambda}\left(\sqrt{x^2+y^2+f_2^2}-f_2\right).
    \label{eq:phase1}
\end{align}

\begin{figure}[tb]
\centering\includegraphics{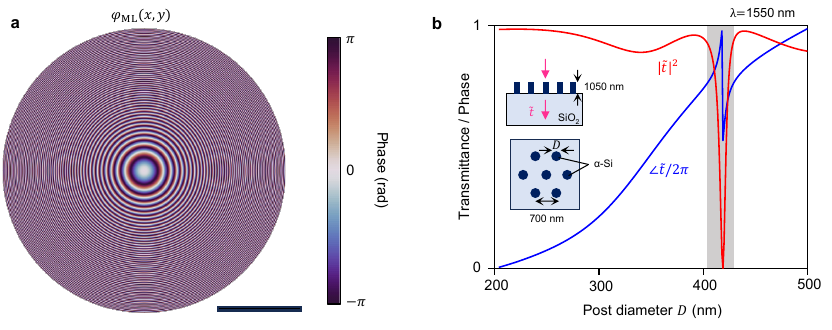}
\caption{
\textbf{ML design for an IM-DD receiver.} 
\figlabel{a} Required phase profile $\varphi_\mathrm{ML}(x,y)$ to realize the ML of the IM-DD receiver in Fig.~\ref{fig:ML-PD}. Scale bar, 100~\textmu m.
\figlabel{b} Simulated transmittance $|\tilde{t}|^2$ and phase $\angle \tilde{t}/2\pi$ for a periodic Si nanopost array on a triangular lattice with a lattice constant of 700~nm (inset). The shaded region is excluded in the ML design to maintain high transmission. 
}
\label{fig:ML-design}
\end{figure}

In this work, we set $\lambda=1550$~nm, $n_s=1.53$, and $f_2 = 525$ {\textmu}m, which is the thickness of the SiO$_2$ substrate. For each device, we adjusted $f_1$ so that the MFD at the PD plane $w_\mathrm{PD}$ becomes $1/\sqrt{2}$ times the PD diameter $D_\mathrm{PD}$. 
$w_\mathrm{PD}$ is given as $w_\mathrm{PD} = w_\mathrm{fiber}(f_1/f_2)/n_s$, where $w_\mathrm{fiber}$ ($= 10.4$~\textmu m) is the MFD of the input fiber.
The phase profile $\varphi_\mathrm{ML}$ used to design the ML of the receiver in Fig.~\ref{fig:ML-PD} is depicted in Fig.~\ref{fig:ML-design}a.

Next, to determine the geometrical shapes of the meta-atoms to achieve $\varphi_\mathrm{ML}(x,y)$, we simulated the transmission characteristics of Si nanoposts using the rigorous coupled-wave analysis (RCWA) method \cite{Liu2012-kd}. Namely, we calculated the complex amplitude $\tilde{t}$ of light transmitted through a periodic array of 1050-nm-high circular Si nanoposts placed on a triangular lattice with a lattice constant of 700~nm.
Simulated transmittance and phase properties ($|\tilde{t}|^2$ and $\angle {\tilde{t}}$) are plotted as a function of the meta-atom diameter $D$ in Fig.~\ref{fig:ML-design}b. We can confirm that an arbitrary phase change ($0\sim2\pi$) can be obtained with a sufficiently high transmittance ($>0.89$) by judiciously selecting $D$ from 204 to 499~nm. 
We could, therefore, derive the shape of the meta-atom at each position $(x,y)$ by one-to-one mapping from $\varphi_\mathrm{ML}(x,y)$ to $D(x,y)$ using Fig.~\ref{fig:ML-design}b.

\subsubsection*{MS for SVR}
\noindent
The polarization-sorting MS used in our SVR was designed based on our previous work \cite{Soma2024-fc}. The Jones matrix of the meta-atom used in this work, which is non-chiral and ideally lossless, can be expressed in a general form as \cite{Arbabi2015-rv, Balthasar_Mueller2017-ex}
\begin{equation}
    \tilde{\mathbf{J}}_\mathrm{MA}=\mathbf{R}(\theta)\begin{pmatrix}

e^{i \varphi_u} & 0 \\
0 & e^{i \varphi_v}
\end{pmatrix}
\mathbf{R}(-\theta),
\label{eq:J_MA}
\end{equation}
where $\varphi_u$ and $\varphi_v$ represent the phase shifts for the eigenmode polarized along the fast and slow axes of the meta-atom, respectively, and $\mathbf{R}(\theta)$ is a rotation matrix with an angle $\theta$ of the meta-atom. From Eq.~(\ref{eq:J_MA}), we can understand that an arbitrary symmetric unitary matrix can be realized by judiciously designing three meta-atom parameters $(\varphi_u,\varphi_v,\theta)$.

\begin{figure}[tb]
\centering\includegraphics{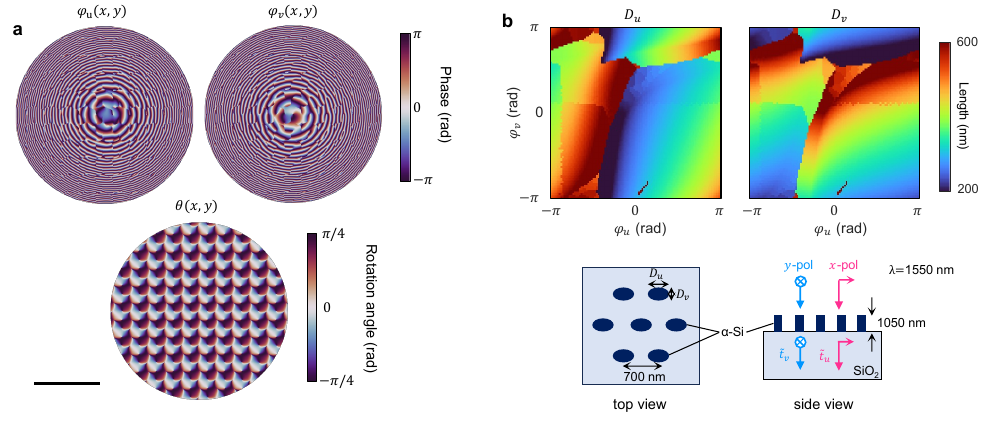}
\caption{
\textbf{MS design for the SVR.} 
\figlabel{a} Spatial distributions of the meta-atom parameters $\varphi_u(x,y)$, $\varphi_v(x,y)$, and $\theta(x,y)$ for the SVR described in Fig.~\ref{fig:SVR}. Scale bar, 100~\textmu m.
\figlabel{b} Calculated dimensions $(D_u,\,D_v)$ of elliptical Si nanoposts that provide phase shifts of $(\varphi_u,\,\varphi_v)$ to the $x$ and $y$-polarized transmitted light. The dimensions were selected from 200 to 600 nm for ease of fabrication. We employed the RCWA method and assumed a periodic array arranged on a triangular lattice with a lattice constant of 700~nm and $\theta = 0$.
}
\label{fig:SVR-design}
\end{figure}

We analytically derived $\tilde{\mathbf{J}}_\mathrm{MA}$ of each meta-atom at $(x,y)$ to realize the polarization-sorting and focusing functionalities to four PDs (see Ref.~\cite{Soma2024-fc} for detailed derivations).
The obtained distributions of three parameters, $\varphi_u(x,y)$, $\varphi_v(x,y)$, and $\theta(x,y)$, of $\tilde{\mathbf{J}}_\mathrm{MA}$ are depicted in Extended Data Fig.~\ref{fig:SVR-design}a. Then, using the one-to-one mapping tables given in Extended Data Fig.~\ref{fig:SVR-design}b, which were generated by the RCWA simulations, $\varphi_u(x,y)$ and $\varphi_v(x,y)$ were mapped to the meta-atom dimensions $D_u(x,y)$ and $D_v(x,y)$ \cite{Arbabi2015-rv, Soma2023-fx}. Here, we assumed Si elliptical nanoposts arranged on a triangular lattice with a lattice constant of 700~nm.

\subsubsection*{MS for CR}
\noindent
For the MS used in our CR, two interleaved sections, MS-A and MS-B, were designed independently.
Here, MS-A and MS-B, comprised of $20\times20$ meta-atoms (see Fig.~\ref{fig:CR}a), were designed to function as polarization splitters for $\pm 45^\circ$ linear ($a$/$b$) and circular ($r$/$l$) polarization bases, respectively. 
The phase profiles $\varphi_p(x,y)$ ($p=a,b,r,l$) required to focus the input lightwave from a fiber to the center position $(x_p, y_p)$ of the corresponding PD are expressed as
\begin{align}
    \varphi_p&(x,y) = \varphi_\mathrm{col}(x,y) -\frac{2n_s\pi}{\lambda}\left(\sqrt{(x-x_p)^2+(y-y_p)^2+f_2^2}-f_2 \right).
\end{align}
For MS-A, the three parameters of each meta-atom were determined in a straightforward manner as $\varphi_{u}=\varphi_{a}$, $\varphi_{v}=\varphi_{b}$, and $\theta=\pi/4$.
On the other hand, the parameters for MS-B were obtained by $\varphi_{u}=(\varphi_{r}+\varphi_{l})/2$, $\varphi_{v}=(\varphi_{r}+\varphi_{l})/2+\pi$, $\theta=(\varphi_{r}-\varphi_{l})/4$ \cite{Arbabi2018-vz}.
We then derived the shapes of all meta-atoms using the one-to-one mapping tables, similar to the design procedure for SVR. In this case, we employed a periodic nanopost array on a square lattice with a lattice constant of 700~nm.

\subsection*{Bandwidth measurement}
\noindent
The setup is shown in the inset of Fig.~\ref{fig:ML-PD}e. The transmitter consisted of a tunable laser source (TLS) and a Mach-Zehnder modulator (MZM) (MX70G, Thorlabs), which was driven by an electrical signal from port 1 of a vector network analyzer (VNA) (MS4647B, Anritsu). The modulated signal at 1550~nm was received by our fabricated device, which was biased through a bias-tee (BT45R, SHF). The AC-coupled photocurrent signal from the membrane PD was sent to port 2 of the VNA.

\subsection*{High-speed signal detection experiments}
\subsubsection*{IM-DD experiment}
\begin{figure}[tb]
\centering\includegraphics{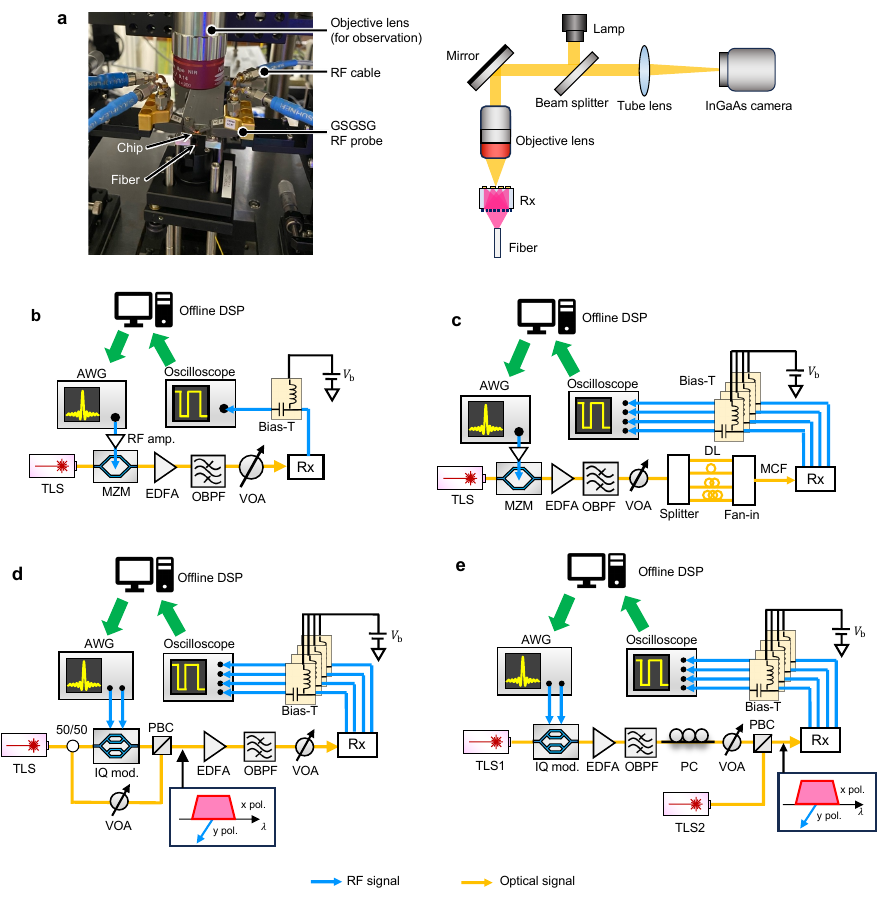}
\caption{
\textbf{Experimental setups.} 
\figlabel{a} Photograph of the basic measurement setup (Left). The two GSGSG RF probes (ACP40-D-GSGSG-250, FormFactor) were used to collect up to four high-speed signals. The right panel shows the schematic of the setup used to observe the positions of the chip, probes, and incident beam.
\figlabel{b-e} Schematics of the setups for the high-speed signal detection experiments for (\textbf{b}) the single-channel IM-DD receiver, (\textbf{c}) four-channel IM-DD receiver, (\textbf{d}) self-coherent detection using the SVR, and (\textbf{e}) coherent detection using the CR.
}
\label{fig:setup}
\end{figure}

\noindent
The experimental setups are shown in Extended Data Figs.~\ref{fig:setup}b,c.
CW light from a TLS (TSL-510, Santac) was intensity-modulated by an MZM, which was driven by a Nyquist-filtered electrical signal from an arbitrary waveform generator (AWG) (M8196A, Keysight: 32~GHz, 92~GS/s). The modulated optical signal was amplified by an erbium-doped fiber amplifier (EDFA) and filtered by an optical bandpass filter (OBPF).
The input optical power to the receiver was controlled by a variable optical attenuator (VOA). 
The membrane PDs on the fabricated devices were DC-biased at $V_b=-3$~V through bias-tees. The electrical signals from the PDs were captured by a real-time oscilloscope (UXR0204A, Keysight: 20~GHz, 128~GS/s). We applied DSP-based equalization to the signals to compensate for the effect of the bandwidth limits of the measurement system, including those of the MZM and RF components. The number of filter taps was chosen to be five and nine for NRZ and PAM4 signal formats, respectively.

For the IM-DD experiment using an MCF (Extended Data Fig.~\ref{fig:setup}c), the optical signal after the VOA was split into four channels by a fiber coupler. They were transmitted through fiber DLs with different lengths so that the signals in different MCF cores were uncorrelated. Four-channel photocurrent signals from the receiver were simultaneously captured by the real-time oscilloscope and demodulated through DSP.

\subsubsection*{Self-coherent experiment using SVR}
\noindent
The setup is shown in Extended Data Fig.~\ref{fig:setup}d.
CW light from the laser (TSL-510, Santac) was split by a 50:50 fiber coupler into two paths to generate a self-coherent signal. The light in the signal path was modulated using a lithium-niobate-based IQ modulator (Ftm7962ep, Fujitsu) driven by two-channel Nyquist-filtered signals from the AWG. The CW power in the other path was adjusted by a VOA to the same level as the signal power. 
They were then combined by a PBC, amplified by an EDFA, filtered by an OBPF, and input to the fabricated SVR.
The input optical power was controlled by another VOA. The photocurrent signals from the membrane PDA were captured by the real-time oscilloscope. 
Through offline DSP with $2\times3$ real-valued multi-input-multi-output (MIMO) equalizers \cite{Mori2012-fn, Faruk2013-uu}, polarization changes inside the fiber were removed, and the original IQ signals were retrieved.

\begin{figure}[tb]
\centering\includegraphics{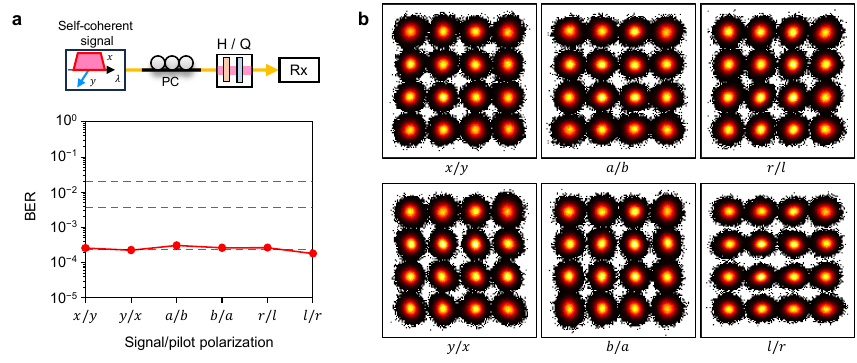}
\caption{
\textbf{Polarization-drift-resilient self-coherent detection using the SVR.} 
\figlabel{a} Measured BER of 160-Gbit/s 16QAM signal at 1550-nm wavelength with different polarization states. Received optical power is $P_\mathrm{in}=-10$~dBm. The top panel shows the schematic of the measurement setup, where we emulate arbitrary polarization drift during the transmission by rotating the HWP (H) and QWP (Q).
\figlabel{b} Retrieved constellation diagrams for different input states of polarization.
}
\label{fig:SVR-pol}
\end{figure}

To emulate arbitrary polarization drift during the transmission, we employed the setup shown in Extended Data Fig.~\ref{fig:SVR-pol}a. 
A polarization controller (PC), composed of HWP and a quarter-wave plate (QWP), was inserted to change the polarization state of the self-coherent signal input to the device.

\subsubsection*{Coherent detection experiment}
\noindent
The experimental setup is shown in Extended Data Fig.~\ref{fig:setup}e.
CW light from a narrow-linewidth laser (TLG-210, Alnair Labs) was modulated by the IQ modulator, driven by two-channel Nyquist-filtered signals from the AWG. The modulated coherent optical signal was amplified by an EDFA, filtered by an OBPF, and its power was adjusted by a VOA.
Then, a LO light from another laser was combined by a PBC and input to the fabricated device. Here, the incident signal and LO light were fixed to the $x$ and $y$ polarization states, respectively.  The photocurrent signals from four PDs were captured by the real-time four-channel oscilloscope. Through offline DSP, the differences between the signals from $a$/$b$ and $r$/$l$ PDs were derived, 
followed by a $2\times2$ adaptive MIMO filter for equalization and retrieval of original IQ signals.

\section*{Acknowledgment}
\noindent
This work was obtained in part from the commissioned research (JPJ012368C08801, JPJ012368C03601) by National Institute of Information and Communications Technology (NICT) and JSPS KAKENHI (JP23H00172, JP24KJ0557).
A part of the device fabrication and characterization was conducted at the cleanroom facilities of d.lab in the University of Tokyo, supported by the "Nanotechnology Platform Program" of the Ministry of Education, Culture, Sports, Science and Technology (MEXT), Japan (JPMXP1224UT1115, JPMXP1223UT0179).
The authors thank K. Misumi, M. Hino, and H. Miyano for their support in device fabrication.
G.S. acknowledges financial support from Japan Society of Promotion of Science (JSPS) 
and World-leading Innovative Graduate Study Program - Quantum Science and Technology Program (WINGS-QSTEP), the University of Tokyo.

\section*{Author contributions}
\noindent
G.S. and T.T. conceived the experiment.
G.S. performed device design, fabrication, measurement, and data analysis.
T.A. performed the wafer bonding process.
E.K. performed the MOVPE crystal growth.
K.K. assisted with the MS fabrication. 
M.T. contributed to the discussion of the bonding process and epitaxial layer design and provided experimental facilities.
Y.N. contributed to the overall discussion and provided experimental facilities.
G.S. and T.T. wrote the manuscript with inputs from all authors.
T.T. supervised the project.

\section*{Competing interests}
\noindent
The authors declare no competing interests.

\clearpage

\begin{figure}[tb]
\centering\includegraphics{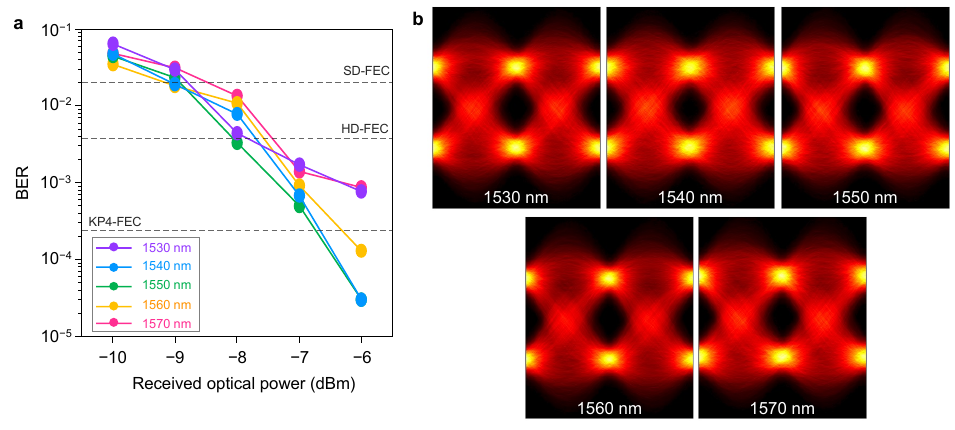}
\caption{
\textbf{Wavelength-independent characteristics of ML-integrated IM-DD receiver across the $C$ band.} 
\figlabel{a} Measured BER of a 40-Gbaud NRZ signal as a function of the received optical power at different wavelengths.
\figlabel{b} Retrieved eye diagrams for all wavelengths.
}
\label{fig:IMDD-C}
\end{figure}

\end{document}